\documentclass[aps,preprintnumbers,prl,twocolumn,superscriptaddress]{revtex4}
\usepackage{epsfig,latexsym,cancel,amssymb,amsmath}
\usepackage{graphicx}
\usepackage{epstopdf}
\usepackage{color}
\usepackage{hyperref}
\usepackage{hhline}
\usepackage[none]{hyphenat}

\usepackage[caption=false]{subfig}
\usepackage{epstopdf}
\usepackage{color}
\usepackage{bm}
\usepackage[normalem]{ulem}
\usepackage{array}
\usepackage{natbib}
\usepackage{booktabs}

\newcommand{\RSN}{R_{\textrm{\tiny{SN}}}}
\newcommand{\RSF}{R_{\textrm{\tiny{SF}}}}
\newcolumntype{A}{ >{\centering\arraybackslash} m{3.5cm} }
\newcolumntype{B}{ >{\centering\arraybackslash} m{2.5cm} }
\newcolumntype{C}{ >{\centering\arraybackslash} m{2cm} }
\newcolumntype{D}{ >{\centering\arraybackslash} m{1.5cm} }
\newcolumntype{E}{ >{\centering\arraybackslash} m{2.5cm} }

\def\beq{\begin{equation}}
\def\eeq{\end{equation}}
\def\bea{\begin{eqnarray}}
\def\eea{\end{eqnarray}}

\begin{document}

\title{Neutrinos from the cosmic noon: a probe of the cosmic star formation history}
\author{Riya}
\affiliation{Department of Mechanical Engineering, Indian Institute of Technology Bombay, Powai, Mumbai 400076, India}
\email{riyasingh@iitb.ac.in}
\author{Vikram Rentala}
\affiliation{Department of Physics, Indian Institute of Technology Bombay, Powai, Mumbai 400076, India}
\email{rentala@phy.iitb.ac.in}

\begin{abstract}
Multiple astrophysical probes of the cosmic star formation history yield widely different inferences of this rate at redshifts $z \gtrsim 1$. While all probes seem to indicate a period of peak star formation known as the cosmic noon between $1.5 \lesssim z \lesssim 3$, the detailed inferences from these probes are in disagreement.  In particular, the magnitude of the peak star formation rate density indicated by H$\alpha$ data is higher by a factor of $\sim$ 4 compared to the magnitude of the peak indicated by UV/IR data. In this work, we explore the potential of future measurements of the diffuse supernova neutrino background at the Hyper-Kamiokande (HK) experiment to resolve the discrepancy and help pin down the magnitude of the peak cosmic star formation rate. We find that, depending upon the cosmic core-collapse supernova neutrino spectrum, HK loaded with 0.1\% Gadolinium by mass has the potential to discriminate between the different star formation histories with between 1.6-20 years of data collection.
\end{abstract}

\maketitle

\section{Introduction}
Understanding the origin and evolution of galaxies is a problem at the frontier of cosmology.  Observations of cosmic structure formation are generally in agreement with predictions of the $\Lambda$CDM cosmology indicated by other cosmological probes such as the cosmic microwave background (CMB)~\cite{Aghanim:2018eyx}. In particular observations of structure on large scales  ($\gtrsim $~10~Mpc) appear to be consistent with purely gravitational dynamics, as seen in $N$-body simulations~\cite{Percival:2013awa,Silk:2016srn, Vogelsberger:2019ynw}. On smaller scales however, the complex physics of galaxies and their evolution has a strong, and at the moment, incompletely understood, effect on cosmic structure. Several problems in small scale structure that appear to be in conflict with standard $\Lambda$CDM cosmology, such as the core-cusp problem~\cite{Oh:2010ea, Rocha:2012jg, Peter:2012jh, 2013MNRAS.431L..20Z} and the too-big-to-fail problem~\cite{BoylanKolchin:2011de,2012MNRAS.423.3740V}, may possibly be resolved by a better understanding of the baryonic physics~\cite{Bullock:2000wn,Benson:2001at,2012MNRAS.422.1231G,DelPopolo:2016emo}.

One of the critical ingredients to understand galaxy evolution is the cosmic star formation history (CSFH). A variety of probes indicate a peak in the global (comoving) star formation rate density (SFRD) between redshifts $1.5\lesssim z \lesssim 3$~\cite{Madau:2014bja}, which suggests that this was the epoch of most vigorous star formation. This epoch is known as the \emph{cosmic noon}. It has been estimated that nearly half of the stellar mass of the universe today was formed during this epoch.

A detailed prediction of the redshift and magnitude of the peak SFRD is lacking~\cite{MacLow1229229}. Qualitatively, it is expected that at early times the star formation rate grows as the dark matter potential wells grow and collect more gas, which in turn cools to form stars. At late times, the densities dilute due to Hubble expansion to the point where collisional cooling of the gas is inhibited, and new star formation in halos is suppressed. Thus, it is between these two extremes that one expects a peak in the global star formation rate~\cite{Hernquist:2002rg}. However, complex feedback processes from galaxies such as stellar radiation, supernova explosions, and radiation from accreting massive black holes, lead to energy deposition in the interstellar medium (ISM) and circumgalactic medium, which would inhibit star formation, delaying the cosmological peak in the SFRD history.

Moreover, there are a large range of galaxy masses (or length scales) that contribute to the cosmic SFRD, the effects of which can only be captured in large-volume cosmological simulations. Modelling the complicated effects of galactic feedback on the ISM in hydrodynamical, large-volume, cosmological simulations (for e.g. \cite{Schaye:2014tpa, Pillepich:2017jle}) is computationally unfeasible in the state-of-the-art simulations at the moment. Thus, simplified prescriptions for star formation efficiency and feedback mechanisms have to be invoked in these simulations using so-called sub-resolution parameters~\cite{Vogelsberger:2019ynw}.

Turning this around, sub-resolution parameters are therefore~\emph{calibrated} by matching hydrodynamical simulations to observations. The SFRD history thus becomes one of the critical calibration observables for galaxy formation simulations~\cite{Vogelsberger:2019ynw}. Therefore, in order to understand galaxy formation and evolution, it is crucial to have a robust understanding of the CSFH from observations. Observations in multiple wavelength regions of the electromagnetic spectrum agree on the approximate redshift range mentioned above for the cosmic noon, however \textit{different probes seem to give differing results for the magnitude of the peak SFRD in this epoch}.  Thus, additional probes of the SFRD from the cosmic noon could help us resolve this discrepancy.

In this paper, we propose to resolve this discrepancy using future observations of the diffuse supernova neutrino background (DSNB).  We will first give a summary of the various conventional probes used to measure the CSFH and highlight the discrepancy in peak star formation rate obtained from UV/IR data as compared to $H\alpha$ data. Next, we will discuss the expected DSNB flux from using CSFHs inferred from UV/IR vs  $H\alpha$ data. After this, we will describe how the Hyper-Kamiokande experiment will measure the spectrum of the DSNB and the limitations set on this measurement from various backgrounds. Finally, we will present our main results on the potential for DSNB measurements at HK to resolve the inferred SFRD discrepancy. Throughout this paper, we use values of cosmological parameters taken from Planck~\cite{Aghanim:2018eyx}: $\Omega_m = 0.31$, $\Omega_\Lambda = 0.69$, and $H_0 = 67$ km s$^{-1}$ Mpc$^{-1}$.

\section{Probes of star formation history}
A comprehensive review of different measurements of the SFRD inferred from various (electromagnetic) probes was made by \citet{Madau:2014bja}, here we will only attempt to summarize the main ideas.

In any standard inference of the star formation rate, mass is inferred from light in different wavelength bands. This conversion requires knowledge of the mass-to-light ratio, which depends on several factors such as galactic age, metallicity, star-formation history, extinction due to dust and initial (stellar) mass function (IMF). Currently, this ratio is obtained using stellar population synthesis models which compute the predicted spectral energy distribution (SED) of a galaxy with a given set of properties~\cite{Conroy:2013if}. Synthesis models evaluate evolutionary tracks, i.e. how luminosity, temperature, and radius of stars vary as functions of mass and age and add in effects of stellar atmospheres to predict the spectra. The spectrum at a given galaxy age is then obtained by summing isochrone spectra of a population of stars over an assumed initial mass function.

\textbf{UV and IR rest-frame emissions:} The starlight from galaxies is typically dominated by young, bright, and massive stars  which emit in the UV in the rest-frame of the galaxy. This UV light suffers extinction from the ISM and can be absorbed and re-emitted as mid- and far-infrared (MIR/FIR) light. Thus, two complementary strategies to study the SFRD are by measuring the rest-frame UV and IR emissions.

At low redshifts  $( z \lesssim 1.4)$, UV emission must be measured using space-based UV instruments. UV emission at larger cosmological distances is redshifted to optical bands, and thus deep surveys with ground based telescopes are a sensitive probe of this emission over a redshift range $1.4 \lesssim z \lesssim 6$.  UV measurements are very sensitive to the assumed IMF, since the light is dominated by the emission from massive stars, but the total mass of the galaxy is dominated  by low mass stars. Moreover, the age, dust extinction and metallicity of a galaxy can all lead to a reddening of the spectrum, making a prediction of the UV emission highly uncertain.

On the other hand, IR emission measurements with satellite based telescopes are sensitive to the SFRD over a lower range of redshifts ($z \lesssim 4$) than UV emission measurements. Inferring the SFRD from IR emission has the advantage of not being affected by further extinction due to dust, but the disadvantage is the complex emission process through which dust grains scatter light in the MIR/FIR range, which makes SED predictions very sensitive to the dust composition and emission models. Also, it is difficult to separate IR emission from star formation processes from IR emission from dust that has been warmed by older stellar populations or active galactic nuclei.

In \citet{Madau:2014bja}, luminosity functions at various redshifts were separately taken from several cosmological surveys that had measured rest-frame far-UV or mid- and far-infrared emissions. For each survey, these luminosity functions were then integrated to obtain co-moving luminosity densities, which were further multiplied by appropriate conversion factors, which were calibrated assuming a Salpeter IMF, to obtain the inferred SFRDs at a given redshift. The CSFH measured using UV data was then corrected for estimated dust attenuation based on the observed reddening of the rest-frame UV spectrum. The CSFH obtained from UV data after this dust correction was found to be in agreement with the CSFH obtained from IR data.

~\citet{Madau:2014bja} also provided a best-fitting function to the combined UV and IR data for the SFRD as a function of redshift as below:
\begin{align}
    \label{eq:UVIRSFR}
        \RSF^{UV/IR}(z) = 0.015 \frac{(1+z)^{2.7}}{1+\big[\frac{(1+z)}{2.9}\big]^{5.6}} \, M_{\small{\odot}} \textrm{ year}^{-1} \textrm{ Mpc}^{-3}.
\end{align}

\textbf{H$\alpha$ emission:}
UV radiation from massive $(M \gtrsim 17 \, M_\odot)$, young OB stars leads to photoionization and excitation of their surrounding gas nebulae. Recombination lines of hydrogen in H\,{\sc ii} regions, such as H$\alpha$ and Ly$\alpha$, can be used to measure the photoionization rate and hence probe the massive star formation rate. Among the nebular SFRD tracers, H$\alpha$ lines in particular, are regarded as one of the most reliable and accessible tracers even in highly obscured star-forming galaxies~(see for e.g.~\cite{2006ApJ...642..775M}). Wide-field, deep, narrow-band surveys are needed to use H$\alpha$ as a probe of the CSFH. Currently, H$\alpha$ probes have measured the CSFH out to a redshift of $z \lesssim 3$.  Future observations with NASA's Nancy Grace Roman Space Telescope (formerly WFIRST-AFTA)~\cite{2015arXiv150303757S,2012arXiv1208.4012G} and ESA's EUCLID mission~\cite{2011arXiv1110.3193L} will give an even clearer picture of the H$\alpha$ luminosity function at high redshifts and the inferred CSFH, especially around the cosmic noon~\cite{Pozzetti:2016cch}. The upcoming James Webb Space Telescope~\cite{Gardner:2006ky} will potentially enable the use of H$\alpha$ SFRD measurements up to a redshift of 6 and possibly beyond.

A major limitation of using H$\alpha$ emission as an SFRD probe is that extinction factors for H$\alpha$ need to be modelled or calibrated by comparing H$\alpha$ inferred SFRDs to FIR determined SFRDs or by using relative line strengths of O\,{\sc ii} and H$\alpha$ (see for e.g.  \cite{ 2013MNRAS.428.1128S}). Additionally, active galactic nuclei (AGNs) can also cause ionization and thus lead to increased H$\alpha$ emission. If this contribution is not accounted for, the SFRDs inferred from H$\alpha$ may be overestimated.

We take the inferred SFRDs at multiple redshifts extracted from H$\alpha$ data from~\citep{Gallego:1995ib, 1998ApJ...495..691T, 2003ApJ...586L.115F, 2003AnA...402...65H, 2004ApJ...615..209H, 2006ApJ...651..142H, Shioya:2007kx, 2007ApJ...657..738L, 2008MNRAS.388.1473G, 2008PASJ...60.1219M, 2009ApJ...696..785S, Ly_2010, 2011PASJ...63S.437T, Gunawardhana:2013fha, 2013MNRAS.428.1128S, 2015MNRAS.453..242S, Sobral:2015tla, 2017MNRAS.471..629M, 2018PASJ...70S..17H, Coughlin_2018, 2020MNRAS.tmp..171K} and fit them to get a benchmark CSFH. Almost all studies assume a Salpeter IMF, use the mass-to-luminosity calibration factors from~\citet{1998ARA&A..36..189K}, and a standard extinction assumption of $A = 1$ mag at all redshifts. However, different studies assume different amounts of AGN corrections to the SFRD, mostly in the range of 10-20\%. We have not attempted to recalibrate the extracted SFRDs from these studies by standardizing the extinction factors or AGN contributions.

We then fit the CSFH from H$\alpha$ data using the broken power-law parametrization of Horiuchi et al.~\cite{2009PhRvD..79h3013H},
\begin{align}
    \label{eq:HalphaSFR}
        \RSF^\textrm{H$\alpha$}(z) = \dot{\rho}_0 \left[(1+z)^{\alpha \eta} + \left( \frac{1+z}{B} \right)^{\beta \eta} +  \left( \frac{1+z}{C} \right)^{\gamma \eta} \right]^{1/\eta},
\end{align}
where $\dot{\rho}_0$ has dimensions of $ M_{\small{\odot}} \textrm{ year}^{-1} \textrm{ Mpc}^{-3}$,  $\alpha$, $\beta$ and $\gamma$ are the power-law exponents in the low, mid and high redshift regimes respectively and $\eta=-10$ is a smoothing parameter.
The constants $B$ and $C$ can be alternatively expressed in terms of constants $z_1$ and $z_2$ (which represent the transition redshifts between the low-mid and mid-high regimes) as, $B=(1+z_1)^{1-\alpha/\beta}$ and $C=(1+z_1)^{(\beta-\alpha)/\gamma}(1+z_2)^{1-\beta/\gamma}$.

Since the H$\alpha$ data does not go to very high redshift, we fix the power-law exponent $\gamma=-2.9$, where the value of the exponent is obtained from the UV/IR data fit in eq.~\ref{eq:UVIRSFR} at high redshifts. We then performed a constrained least-squares fit by choosing suitable ranges of the other fitting parameters, and we find the following best-fit values: $\dot{\rho}_0 = 0.01$, $\alpha= 3.95$, $\beta =6$, $z_1 = 0.42$, $z_2 = 1.3$. We have not considered the redshift errors on the H$\alpha$ data when performing our fit.

\begin{figure}
    \begin{center}
    \includegraphics[width=7.5cm]{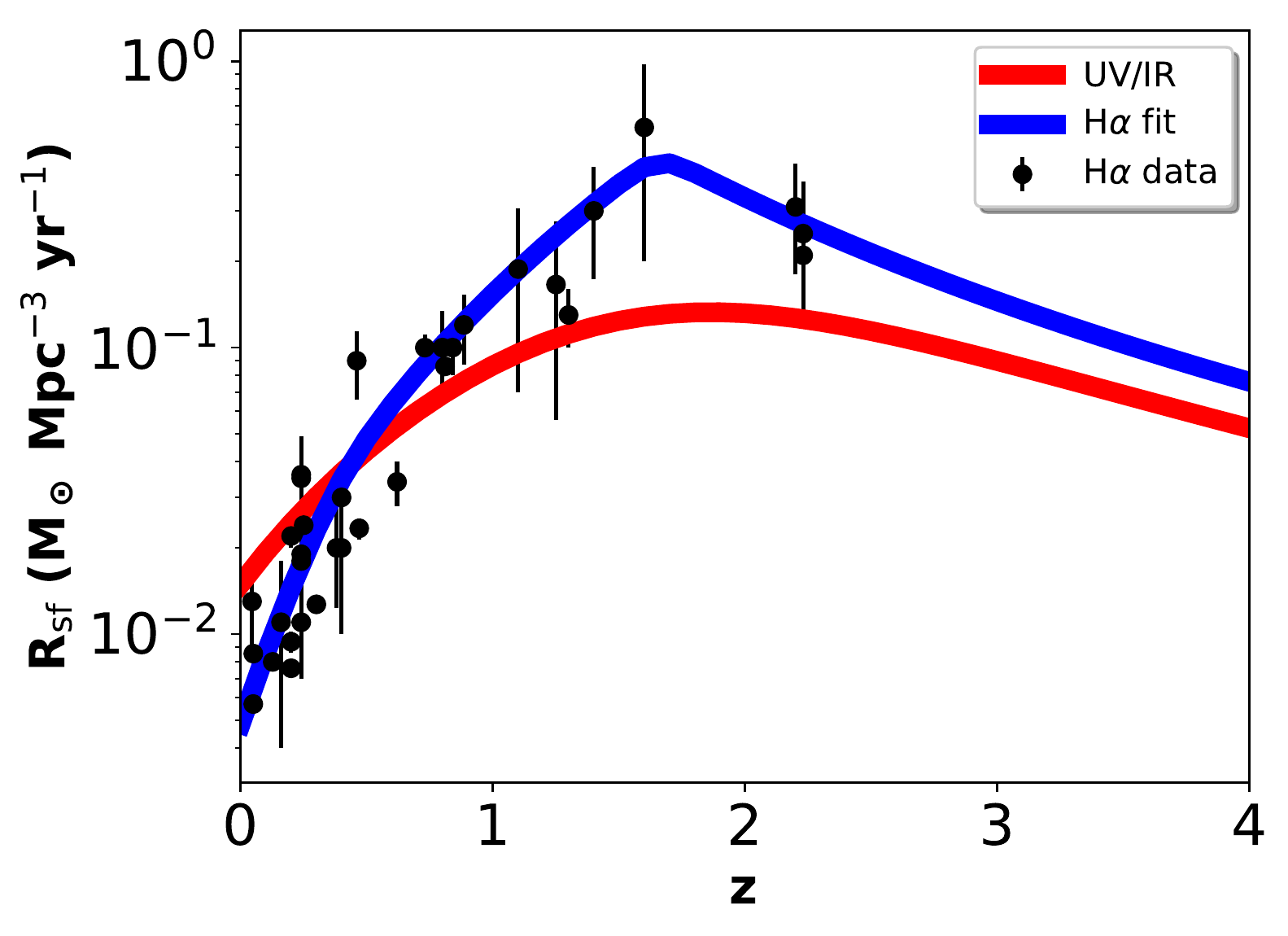}
    \end{center}
    \caption{Cosmic star formation history: We show a comparison between the fit to the CSFH from UV/IR data eq.~\ref{eq:UVIRSFR} (red curve) and our benchmark fit to the $H\alpha$ data eq.~\ref{eq:HalphaSFR}  (blue curve). The inferred SFRDs from H$\alpha$ data with error bars (black data points) are also shown. At low-redshifts, the SFRD, as inferred from $H\alpha$ data, is lower than the SFRD seen from UV/IR data. However, there is also a clear preference for a larger SFRD at cosmic noon preferred by the $H\alpha$ data. }
    \label{fig:SFR}
\end{figure}

The final inferred CSFH from H$\alpha$ data and our best-fit function with the form of eq.~\ref{eq:HalphaSFR} is shown in fig.~\ref{fig:SFR}.  This is to be compared against the CSFH of~\citet{Madau:2014bja} in eq.~\ref{eq:UVIRSFR} using UV/IR emission data which is also plotted in the same figure.

Our fit shows a clear difference between the CSFH inferred from UV/IR data and from H$\alpha$ data. At low redshifts, H$\alpha$ data indicates a lower SFRD than that inferred from UV/IR data. It is also interesting to see that both the H$\alpha$  and UV/IR data indicate a peak in the CSFH around $z\sim 1.8$, i.e. the cosmic noon, however, our H$\alpha$ fit indicates a higher peak SFRD by a factor of $\sim 4$. The discrepancy between H$\alpha$ and UV data at low redshifts and in particular in the local universe has already been noted in the literature (see for e.g. ref.~\cite{Lee:2009by} and references therein). The collected data of \citet{2013ApJ...770...57B} also shows a similar discrepancy between the H$\alpha$ data and UV/IR data at $z\sim2$.

It is possible that calibration uncertainties on the conversion factors from luminosity to SFRD and their redshift dependence could explain the apparent difference between UV/IR data and H$\alpha$ data (see for e.g.~\cite{2019MNRAS.490.5359W}). Other uncertainties such as underestimated UV extinction factors near cosmic noon and an underestimated AGN contribution to H$\alpha$ at high redshifts could also explain the difference between the inferred CSFHs.

Nevertheless, the discrepancy between the CSFHs inferred from UV/IR data and H$\alpha$ data, if taken at face value, motivates us to ask if there are additional, independent probes of the SFRD that could help resolve this discrepancy.

\subsection{Other probes of the CSFH}
Other probes to measure the CSFH include rest-frame radio emission from electrons in H\,{\sc ii} regions (see for e.g.~\cite{2020ApJ...888...61M}) and X-ray emission from newly formed X-ray binaries~(see references in~\cite{Madau:2014bja}). However, the radio emission signal is expected to be weaker from high redshifts where electrons lose energy due to scattering with the higher energy CMB photons at these redshifts. Moreover, calibration is difficult because of the unknown supernova and cosmic ray production rates which can inject free electrons into the gas, as well as the uncertain synchrotron efficiency. X-ray emission from AGNs dominates the cosmic X-ray background making X-ray observations also a less reliable tool to study the CSFH.

Another possible probe of the CSFH is to use the observed rate of core-collapse supernova (CC SNe)~\cite{2011ApJ...738..154H,2012ApJ...757...70D}. This assumes that massive star formation and death is an approximately equilibrium process~\footnote{One reason why this is a good approximation is that the lifetime of massive stars is about 10 million years, which is very short compared to cosmological timescales.}. Unfortunately, poor statistics and dust-obscuration make the observed rate of CC SNe using emitted light unreliable as an SFRD indicator~\cite{Horiuchi:2011zz}. Another issue with using light to track the supernova rate is that many collapses may be intrinsically optically dark or dim~\cite{Lien:2010yb,Horiuchi:2011zz}.

However, 99\% of the energy emission from a CC SNe is expected to be in the form of neutrinos. Neutrinos from CC SNe form the diffuse supernova neutrino background, which is detectable in terrestrial neutrino detectors~\cite{2009PhRvD..79h3013H}. Using weakly-interacting neutrinos as a probe of the CC SNe rate has the advantage that the detectors have full-sky coverage automatically and neutrinos do not suffer dust obscuration. The downside is that the detection efficiency of neutrinos is low, which makes a measurement of the DSNB and consequent inferences of the CSFH difficult.

We will take the UV/IR CSFH of eq.~\ref{eq:UVIRSFR} and our best-fit CSFH to H$\alpha$ data parameterized by eq.~\ref{eq:HalphaSFR} (both of which are shown in fig.~\ref{fig:SFR}) as reference hypotheses for the CSFH. In the rest of this paper, we will explore the ability of future detection and measurement of the spectrum of the DSNB to resolve the difference between these CSFHs by ruling out one hypothesis in favor of another.

\section{DSNB as a probe of the CSFH}
The diffuse supernova neutrino background (DSNB) is the flux of neutrinos and anti-neutrinos from all CC SNe in the causally-reachable universe. The DSNB has yet to be discovered, although efforts are underway already at the Super-Kamiokande (SK) detector in Japan~\cite{Bays:2011si}, and its proposed successor Hyper-Kamiokande~(HK)~\cite{Abe:2018uyc}.
These detectors primarily search for electron anti-neutrinos ($\Bar{\nu}_e$) in the DSNB  by looking for Cherenkov light from positrons created through the inverse $\beta$-decay processes ($\Bar{\nu}_e + p \rightarrow e^+ + n$). While a discovery of the DSNB may take only 5 years at the upgraded Super-Kamiokande gadolinium (SK-Gd) detector~\cite{Horiuchi:2008jz,Sekiya:2020pun}, characterization of the full spectrum of the DSNB, which is needed for extraction of the CSFH, would need a larger volume detector such as HK and will take longer, possibly 10 - 20 years.

The expected DSNB flux spectrum at Earth is given by~\cite{Beacom:2010kk},
\begin{align} \label{eq:DSNBflux}
    \frac{d\Phi}{dE_\nu} = \int_{0}^{\infty} \phi_{\textrm{\tiny{SN}}} [E_\nu (1+z)]  \RSN(z)  (1+z)   \left |c\frac{dt}{dz}\right |dz .
\end{align}
Here, $\frac{d\Phi}{dE_\nu}$ is the differential arrival flux of neutrinos with incident energy  $E_\nu$. The right-hand side of this equation relates the arrival flux to the emission spectrum  $\phi_{\textrm{\tiny{SN}}}$ of neutrinos from a single ``cosmologically-averaged'' CC~SN (where the arriving neutrino had an energy $E_\nu (1+z)$ at the time of emission) and the supernova rate density $\RSN$ (measured in number/Mpc$^3$/yr) at redshift $z$. The redshift-time conversion factor over the redshift range of interest is given as usual by,
\begin{align}
 \label{eq:redshifttime}
    \left | \frac{dt}{dz} \right | =\frac{1}{H_0(1+z)}\frac{1}{\sqrt{\Omega_m(1+z)^3 + \Omega_\Lambda}}
\end{align}
for a $\Lambda$CDM universe.

The supernova rate can then be related to the star formation rate density at a given redshift by the relation~\cite{Beacom:2010kk},
\begin{equation}
    \RSN(z) =\frac{ \int_8^{100} dM \left( \frac{dN}{dM} \right )}{\int_{0.08}^{100}  dM \left( M \frac{dN}{dM}\right )}\RSF(z)=  \frac{\RSF(z)}{147M_\odot}, \label{eq:SNRSFRrelation}
\end{equation}
where $\frac{dN}{dM}$ is the IMF, and we have made the assumption that all stars above $8\,M_\odot$  will end as CC~SNe. To be consistent with the extracted SFRD assumptions in the previous sections, we have assumed a Salpeter IMF with $dN/dM \propto M^{-2.35}$ and a minimum and maximum stellar mass of $0.08 \, M_\odot$ and $100\,M_\odot$, respectively.

\subsection{Supernova Neutrino Emission Spectrum}
At the moment, we have a very limited understanding of the precise neutrino spectra emitted from a CC~SN, or the diversity of neutrino spectra that are possible from different CC~SNe. The only observational probe that we have is from SN1987A in the Large Magellanic Cloud, with $\mathcal{O}(10)$ neutrinos each seen by the Kamiokande~\cite{PhysRevLett.58.1490,PhysRevD.38.448},  IMB~\cite{PhysRevLett.58.1494,PhysRevD.37.3361} and Baskan~\cite{1987JETPL..45..589A} detectors. While a CC~SN in the Milky Way would generate several thousand neutrino-detection signals in modern detectors such as the 32 kton SK experiment~\cite{Ikeda_2007} or DUNE~\cite{Migenda:2018ljh}, the rate for galactic SNe is very low $\sim$~2.5 per century~\cite{1994ApJS...92..487T}.

It has been proposed in ref.~\cite{PhysRevLett.95.171101} to measure the neutrinos from CC SNe in the nearby universe (within $\sim$ 10 Mpc) with a Mton scale detector, as a faster method for calibrating the CC SNe spectrum. With future Mton scale detectors such as Hyper-Kamiokande~\cite{2003IJMPA..18.4053N}, UNO~\cite{Jung:1999jq}, and MEMPHYS~\cite{deBellefon:2006vq,Agostino:2012fd}, we would be able to see $\sim$ 1 neutrino per supernova, but the SN rate in this enlarged neighbourhood would also be larger $\sim$ 1/year. A more recent, innovative proposal to look for CC SNe in our galactic neighborhood using archaeological lead-based cryogenic detectors to detect coherent, elastic neutrino scattering was suggested in ref.~\cite{Pattavina:2020cqc}.

A detector mass scale of $\sim$~5-10 Megatons (Deep-TITAND~\cite{Suzuki:2001rb}, MICA~\cite{Boser:2013oaa}) would permit observations of mini-bursts of neutrinos from CC SNe in nearby galaxies on a roughly yearly basis~\cite{Kistler_2011}. For a run of a few decades of such a 5 Mton detector, it is quite probable that an SN occurs in one of the Milky Way, M31, M33, or their smaller satellite galaxies. This would lead to at least one burst with  $\sim$~$10^2$ - $10^6$ events.

These observations would feed in to calibration of numerical models of CC SNe \cite{Ando:2005ka,Yuksel:2005ae,Lunardini:2012ne,Adams:2013ana,Nikrant:2017nya,Horiuchi:2017sku,Migenda:2020rot}, and over the time scale of a decade or two, they could help us predict with much more precision the parameters describing the CC SNe neutrino spectrum. One additional advantage of these measurements is that the data would be averaged over many supernovae, which is useful if the emission from different CC~SNe is less uniform than expected.

In this work, we will assume that the spectrum of CC~SNe will be well characterized by a combination of simulations and possible calibration to observations of nearby SNe neutrinos, and we will neglect any uncertainty in the spectrum. For the rest of this paper, we will assume that the cosmologically-averaged CC~SNe $\Bar{\nu}_e$ spectrum (the type of DSNB neutrinos which are being looked for at SK and HK) is of the Fermi-Dirac form~\cite{Beacom:2010kk},
\begin{equation}
\phi_{\textrm{\tiny{SN}}}(E_\nu) = \frac{E_{\textrm{tot}}}{6} \frac{120}{7\pi^4} \frac{E_\nu^2}{T^4} \frac{1}{e^{E_\nu/T}+1}.
\label{eq:emissionspectra}
\end{equation}
This spectrum represents the effective time-integrated emission from the neutrino burst of a CC~SNe with free parameters $E_{\textrm{tot}}$ and $T$. $E_{\textrm{tot}}$ is the total energy emitted in the CC~SNe explosion and the factor of 6 assumes equal partitioning into all neutrino flavors and in particular to $\Bar{\nu}_e$, and $T$ describes the effective temperature of the $\Bar{\nu}_e$. We assume that the properties of cosmologically-averaged CC~SNe (and hence the parameters $E_{\textrm{tot}}$ and $T$) are redshift independent.

For SN1987A, with the sparse data that was observed, spectral parameters $E_{\textrm{tot}} =  3\times10^{53}$ erg and $T\simeq 5$~MeV were found to be consistent with the observations~\cite{Beacom:2010kk}. Different choices of $E_{\textrm{tot}}$ give rise to different normalizations of the total DSNB flux. An estimate for the range of $E_{\textrm{tot}}$ values based on the range of uncertainty of binding energies of neutron star final states is between $2.4 \times 10^{53}$~erg and $3.5 \times 10^{53}$~erg~\cite{Yoshida:2005uy}. For the rest of this paper, we take the benchmark value of $E_{\textrm{tot}} =  3\times10^{53}$~erg. We will, however, consider several different benchmark temperatures, since the choice of temperature will crucially determine the shape of the spectrum and relative contribution of neutrinos from cosmic noon at a particular neutrino arrival energy. We will consider the benchmark scenarios where $T= 4, 6, 8$~MeV and another benchmark scenario where 70\% of CC~SNe emit neutrinos with an effective temperature $T=4$~MeV and the other 30\% emit with a temperature of 8~MeV. We refer to this latter benchmark as the ``mixed'' case, which represents a situation where a fraction of the DSNB is sourced by failed CC~SNe  (where the stellar collapse ends in direct black hole formation), which have a higher effective temperature~\cite{Lunardini:2009ya,Tamborra:2016lpf, Priya:2017bmm}.

\vspace{5mm}

The location of the peak in the neutrino emission spectrum depends on the effective temperature. The DSNB flux (eq.~\ref{eq:DSNBflux}) involves contributions of neutrinos from many different redshift slices, weighted by the SFRD at the corresponding redshift, where the spectrum from a given redshift slice is shifted in energy by a factor $(1+z)$.

\begin{figure}
    \centering
    \includegraphics[scale = 0.5]{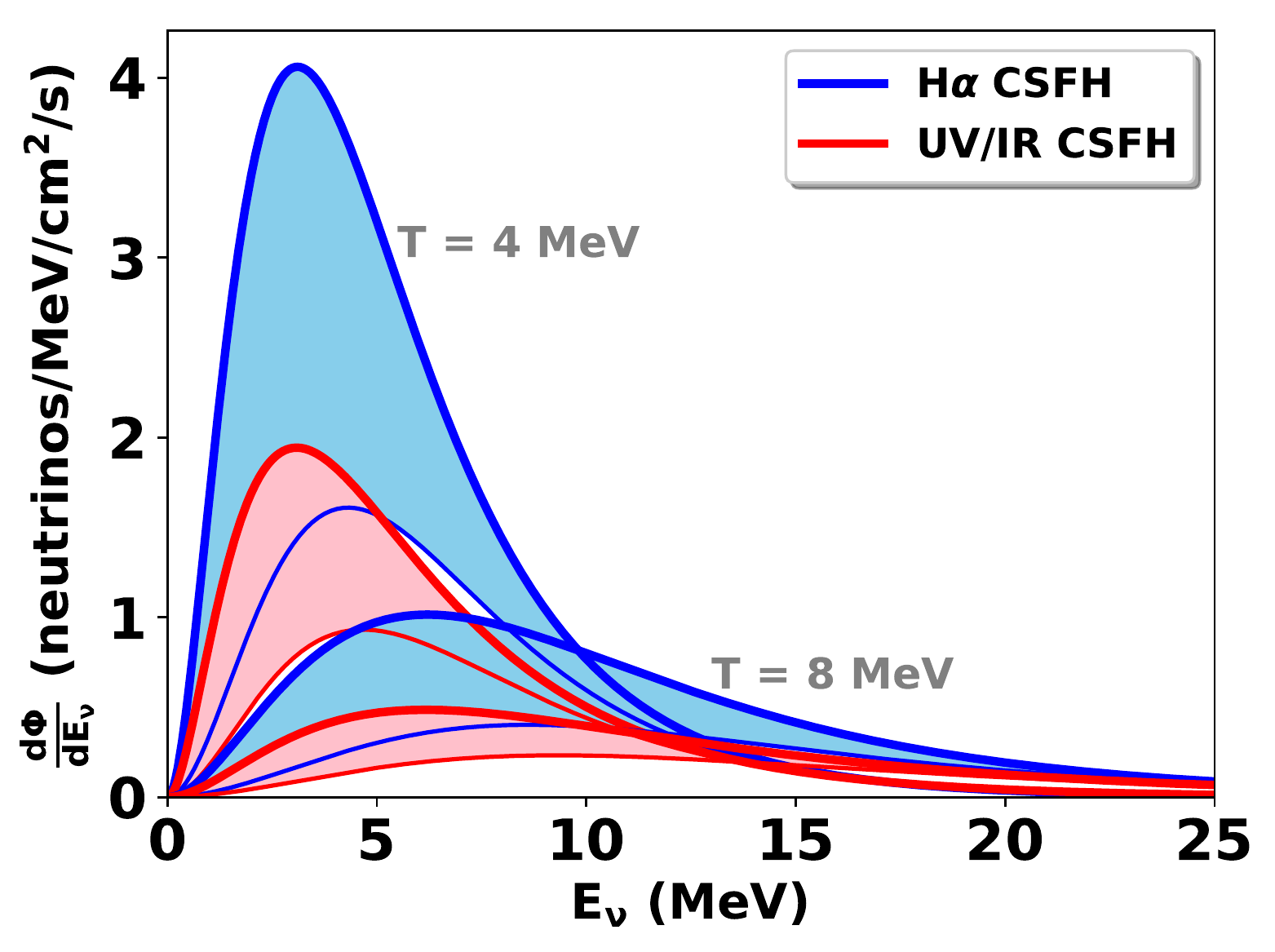}
    \caption{Predicted DSNB flux spectra assuming different CSFHs. The thick red lines show the flux spectrum assuming the UV/IR inferred CSFH and the thick blue lines show the flux spectrum assuming the H$\alpha$ inferred CSFH. We have shown the spectra assuming two different  temperature parameters $T= 4$~MeV (curves with peaks to the left) and 8~MeV (curves with peaks to the right). The shaded bands indicate the contribution to the DSNB flux from cosmic noon, i.e. from $1.5 \lesssim z \lesssim 3.5$. The rest of the flux is dominantly from redshifts  $z\lesssim 1.5$. The difference in SFRDs at cosmic noon inferred from UV/IR data versus H$\alpha$ data greatly accentuates the difference in the predicted DSNB spectra, especially at lower energies. }
    \label{fig:Fluxarr}
\end{figure}

In fig.~\ref{fig:Fluxarr}, we show the predicted DSNB flux as a function of neutrino arrival energy for both the UV/IR CSFH from~eq.~\ref{eq:UVIRSFR} (thick red curve) and H$\alpha$ CSFH from eq.~\ref{eq:HalphaSFR} (thick blue curve) for the benchmark scenarios where  $T=4$~MeV and $T=8$~MeV. Super-Kamiokande has set a DSNB flux limit of 3 neutrinos/cm$^{2}$/s above neutrino energies of 17.3 MeV~\cite{Bays:2011si}. For the $T=8$~MeV case, the integrated DSNB flux that we predict for the UV/IR (H$\alpha$) CSFH for neutrino energies $> 17.3$~MeV is 1.4 (1.9) neutrinos/cm$^2$/s which is consistent with the limit set by SK. Since the neutrino spectra shifts towards lower energies at lower SN temperatures, all our benchmark SN spectra hypotheses are consistent with the SK bound for either choice of the CSFH~\footnote{In their 2003 analysis~\cite{Zhang:1988tv}, SK had a threshold of 19.3~MeV for the neutrino energy and had set a flux exclusion limit of 1.2 neutrinos/cm$^{2}$/s above this energy. We find a flux of 1.09 (1.38)~neutrinos/cm$^{2}$/s above this threshold for the CC SNe temperature $T=8$~MeV for the UV/IR (H$\alpha$) CSFH. However, the 2003  limit was subsequently revised in a 2011 reanalysis~\cite{Bays:2011si} and the limit was weakened to 1.9~neutrinos/cm$^{2}$/s. Both our CSFH hypotheses are consistent with this revised limit.}.

We also show the relative contribution from neutrinos from the cosmic noon in fig.~\ref{fig:Fluxarr}, i.e. from redshifts $1.5 \lesssim z \lesssim 3.5$ using  bands for the two CSFHs. The contribution from neutrinos beyond redshift 3.5 is negligible. From the figure, we can see that the contribution of neutrinos from the cosmic noon is significant at low neutrino energies. Moreover, it can be seen that the difference in the cosmic noon SFRDs inferred from UV/IR versus H$\alpha$ data greatly accentuates the predicted difference in the neutrino spectra from these two CSFHs, especially at lower energies. For $T=8$ MeV, the cosmic noon contributes 55\% (62\%) of the flux at 5~MeV and 23\% (35\%) at 15~MeV for the UV/IR (H$\alpha$) CSFH. At higher energies above 25~MeV, the cosmic noon contributes less than 10 - 15\% of the DSNB flux.

The energy range of sensitivity to DSNB flux measurements at SK and HK are limited by various backgrounds which we will discuss in further detail in a later section. In order for DSNB spectral measurements to be sensitive to the differences in contributions from the cosmic noon for different CSFHs, it is advantageous to go to lower threshold energies.
In their 2011 analysis setting an upper limit on the DSNB, SK had set a lower threshold of 17.3~MeV for neutrino energy which was limited by spallation backgrounds~\cite{Bays:2011si} and would not have been very sensitive to the cosmic noon neutrinos, even if the flux normalization was high enough to permit discovery of the DSNB. In the upgraded SK-Gd detector, it is expected that the threshold might be reduced to as low as $\sim$ 10 MeV~\cite{Sekiya:2020pun} and this could open up sensitivity to differences in neutrinos from the cosmic noon.

For the $T=8$~MeV benchmark spectrum, we find an integrated flux of 3.4 (5.7)~neutrinos/cm$^2$/s for neutrinos above 10~MeV for the UV/IR (H$\alpha$) CSFHs. Moreover, for the integrated flux in this range, 21\% (33\%) of the total neutrinos are from the cosmic noon.

In table~\ref{tab:fluxes}, we give numerical values of the fluxes of neutrinos from the cosmic noon at various energies/energy ranges corresponding to different assumptions of the CC SNe temperature parameter for both the UV/IR and H$\alpha$ CSFHs. In table~\ref{tab:fluxratios}, we show the relative fraction of neutrinos from the cosmic noon at these energies/energy ranges for different assumptions of the CC SNe temperature parameter. We can clearly see from these tables that measurements of the neutrino flux at lower threshold energies will be more sensitive to a higher fraction of the cosmic-noon neutrinos.

\begin{table}
\begin{centering}
\setlength{\tabcolsep}{0.38em}
\begin{tabular}{|c|c|c||c|c|}
  \hline
   \toprule
& \multicolumn{2}{c||}{Energy} & \multicolumn{2}{c|}{Energy Range} \\ \hline
Temp &10 MeV & 17.3 MeV  & $>$ 10 MeV & $>$ 17.3 MeV  \\ \hline
  4 MeV & 0.5 (0.77)  & 0.08 (0.08) & 2.0 (2.5)  & 0.34 (0.29) \\
  6 MeV & 0.5 (0.94) & 0.15 (0.21) & 3.1 (4.6) & 0.92 (1.0) \\
  8 MeV & 0.4 (0.8) & 0.18 (0.29) & 3.4 (5.7) & 1.4 (1.9)   \\
  Mixed & 0.47 (0.78) & 0.11 (0.14) & 2.4 (3.4)  &  0.66 (0.77)\\
  \hline
\end{tabular}
\caption{Table of fluxes from the cosmic noon for given energies/energy ranges (columns) assuming the UV/IR CSFH. The rows correspond to different assumptions of the CC SNe temperature spectral parameter. Numbers in brackets indicate the corresponding fluxes assuming the H$\alpha$ CSFH. The two columns on the left are fluxes at a specific energy and have units of neutrinos/MeV/cm$^2$/s, whereas the two columns on the right are fluxes integrated over the indicated energy range and have units of neutrinos/cm$^2$/s. The flux at lower threshold energies is significantly higher for the H$\alpha$ CSFH, primarily due to the contribution of a larger fraction of neutrinos from the cosmic noon.}
\label{tab:fluxes}
\end{centering}

\end{table}

\begin{table}
\begin{centering}
\setlength{\tabcolsep}{0.38em}
\begin{tabular}{|c|c|c||c|c|}
  \hline
     \toprule
& \multicolumn{2}{c||}{Energy} & \multicolumn{2}{c|}{Energy Range} \\ \hline
Temp &10 MeV & 17.3 MeV  & $>$ 10 MeV & $>$ 17.3 MeV  \\ \hline
  4 MeV & 0.12 (0.23) & 0.01 (0.04) & 0.06 (0.13)  & 0.01 (0.02)\\
  6 MeV & 0.28 (0.40)  & 0.08 (0.17)  & 0.14 (0.25) & 0.04 (0.09) \\
  8 MeV & 0.40 (0.50)  & 0.17 (0.29) & 0.21 (0.33) & 0.08 (0.18) \\
  Mixed & 0.19 (0.31)  & 0.09 (0.19)  & 0.12 (0.23) & 0.05 (0.14) \\
  \hline
\end{tabular}
\caption{Table of DSNB flux fractions arising from the cosmic noon $(1.5\lesssim z \lesssim 3.5)$ assuming the UV/IR CSFH for given energies/energy ranges (columns). The rows correspond to different assumptions of the CC SNe temperature spectral parameter.  Numbers in brackets indicate the corresponding fractions for the H$\alpha$ CSFH. Measurements of the DSNB flux with a lower threshold will receive greater contributions from the cosmic noon and will thus be more sensitive to the differences between the UV/IR and  H$\alpha$ CSFH.}
\label{tab:fluxratios}
\end{centering}
\end{table}

Next, we will discuss the proposed Hyper-Kamiokande experiment and explore whether HK would be sensitive to the difference in predicted neutrino fluxes under the two different hypotheses of UV/IR and H$\alpha$ inferred CSFHs.

\section{DSNB detection at Hyper-Kamiokande}
Hyper-Kamiokande is a next-generation underground water Cherenkov neutrino detector, based on the Super-Kamiokande experiment~\cite{Abe:2018uyc}. Electron anti-neutrinos $\Bar{\nu}_e$, are detected by looking for Cherenkov light from positrons created through inverse $\beta$-decay processes ($\Bar{\nu}_e + p \rightarrow e^+ + n$), where the protons taking part in the inverse $\beta$-decay process are from hydrogen atoms in the water. The proposal for the HK experiment is to build two identical water Cherenkov detectors, each having a fiducial volume of 187 ktons. The first detector will be built in the Tochibora mine in Japan, and the second detector is proposed to be built in Korea~\cite{Abe:2016ero}.

The positron rate spectrum at HK is related to the incident neutrino flux $\Phi$ as~\cite{Beacom:2010kk, Lunardini:2010ab},
\begin{align}
\label{eq:positronspectrum}
     \frac{dN}{dE_{e^+}dt} = \frac{d\Phi}{dE_\nu}\sigma_{\textrm{\tiny{inv}}\beta} N_p \epsilon_{\textrm{\tiny{eff}}}.
 \end{align}
 Here, the positron energy in a given reaction $E_{e^+}$ is related to the incident neutrino energy $E_\nu$ as,
 \begin{equation}
 E_{e^+} \approx  (E_{\nu} - 1.3~\textrm{MeV}) (1- E_\nu/m_p),
 \end{equation}
  where $m_p$ is the proton mass.  The cross-section for the inverse $\beta$-decay process is denoted as $\sigma_{\textrm{\tiny{inv}}\beta}$, the number of reaction target protons is $N_p= 2.5 \times 10^{34}$ with $2 \times 187$ kton tanks of water~\footnote{The SK experiment has a fiducial volume of 22.5~kton, and thus has an order of magnitude less target protons than the HK experiment.} and the detector efficiency is denoted as $\epsilon_{\textrm{\tiny{eff}}}$.

 The cross-section for the inverse $\beta$-decay process for an  incident neutrino energy $E_\nu$ is given by,
\begin{equation}
    \sigma_{\textrm{\tiny{inv}}\beta}=0.0952 \times 10^{-42} \textrm{ cm}^2 \left (E_\nu-1.3 \right)^2 \left(1-\frac{7E_{\nu}}{m_p} \right )
\end{equation}
where $m_p$ and $E_\nu$  are both expressed in MeV. For the $\mathcal{O}(10)$~MeV range of neutrino energies that we are considering, the cross-section grows as $\sim E_\nu^2$, which leads to preferential detection of higher energy neutrinos.

Several sources of background must be rejected by applying cuts to the data, some of which are discussed in the next section. At SK these cuts reduced the efficiency of DSNB signal detection to a level of $\epsilon_{\textrm{\tiny{eff}}}=70-80$\%~\cite{Bays:2011si}.

In order to reject background and identify DSNB signal, it is useful to lower the threshold sensitivity of the detector by attempting to identify the neutron produced in the inverse $\beta$-decay reaction by tagging it.  For the Super-Kamiokande detector, it has been proposed to load gadolinium (Gd) in the detector in a project known as SK-Gd \cite{Beacom:2003nk,Sekiya:2020pun} (formerly GADZOOKS!). The basic idea is that the neutron capture efficiency on gadolinium is high and the subsequent de-excitation of Gd produces 8 MeV gamma cascade photons which inverse compton-scatter off of electrons in the detector and can be seen as a slightly-delayed, isotropic, secondary signal in the detector in addition to the primary prompt positron. Loading SK with 0.1\% Gd by mass has been proposed, which would lead to a 90\% tagging efficiency of genuine inverse $\beta$-decay events, slightly lowering the detector signal efficiency $\epsilon_{\textrm{\tiny{eff}}}$ further. We will work with the assumption of 0.1\% Gd loading at HK, and we take the overall positron signal efficiency to be $\epsilon_{\textrm{\tiny{eff}}} = 70$\% after Gd loading at HK.

\begin{figure}
    \centering
    \includegraphics[scale = 0.5]{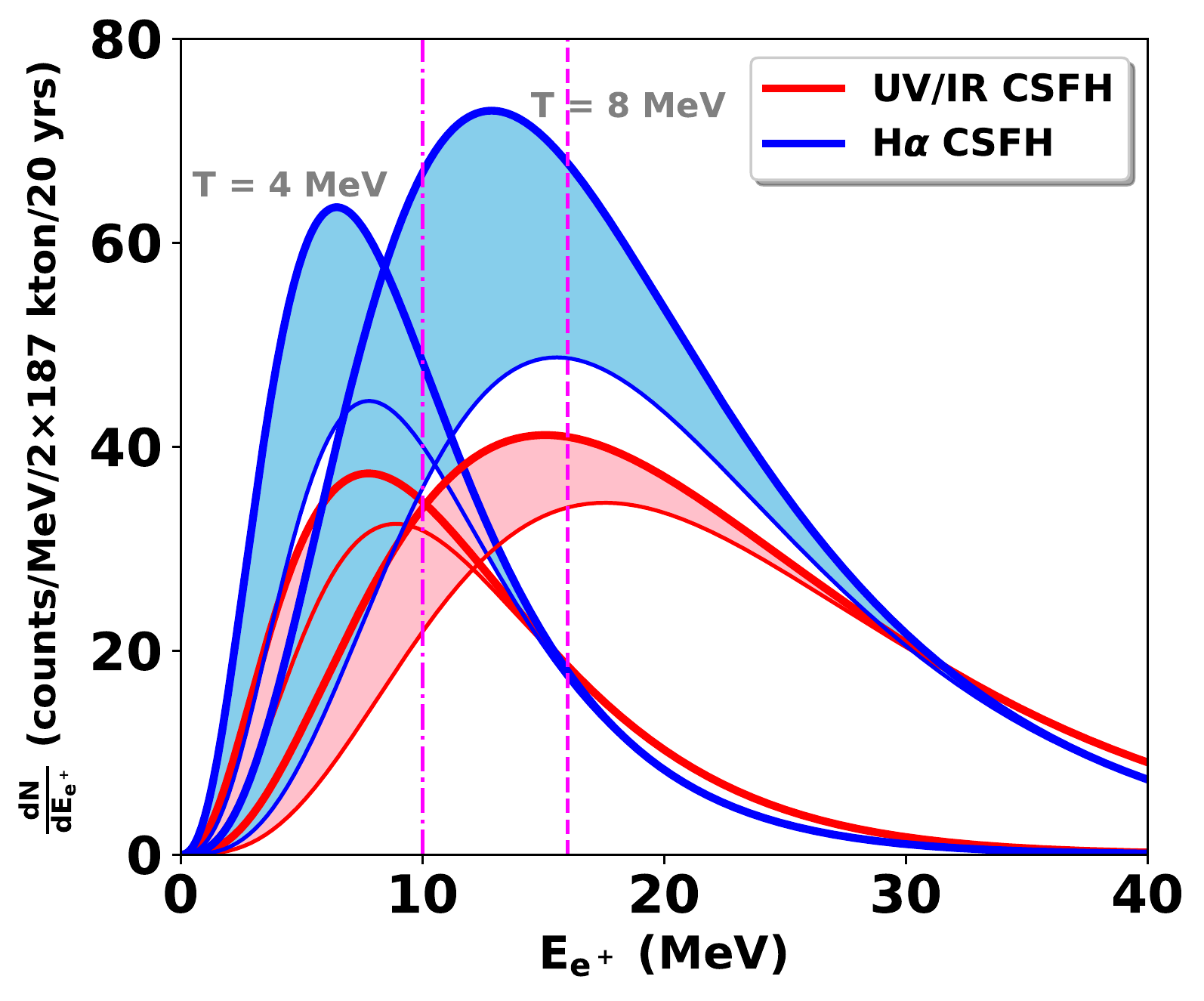}
    \caption{Expected positron spectrum with 20 years of observation at Hyper-Kamiokande with $2\times 187$~kton tanks (fiducial volume) for different assumptions of the CSFH and the CC SNe temperature spectral parameter. The bands indicate contributions to the spectra from cosmic noon neutrinos ($1.5\lesssim z \lesssim 3.5$). We have assumed a 70\% net signal efficiency, taking into account various background reduction cuts on the data and assuming Gd loading at HK to tag accompanying neutrons in inverse $\beta$-decay events. The dashed vertical line indicates the analysis threshold of 16~MeV used in the 2011 DSNB search at Super-Kamiokande without Gd loading, whereas the dot-dashed vertical line indicates a threshold of 10~MeV which is expected with neutron tagging after Gd loading at both SK-Gd and HK.}
    \label{fig:positronspectrum}
\end{figure}
Using eq.~\ref{eq:positronspectrum}, we can now compute the expected positron spectrum under the assumptions of the UV/IR CSFH and the H$\alpha$ CSFH. In fig.~\ref{fig:positronspectrum}, we show the expected positron spectrum at HK with 20 years of data assuming that both detectors are operational, for the scenarios where $T=4$~MeV and $T=8$~MeV. We show the contribution of positrons arising from cosmic noon ($1.5\lesssim z \lesssim 3.5$) neutrinos by using shaded bands in the figure. We also tabulate the positron fractions at a specific energy/energy ranges arising from the cosmic noon in table~\ref{tab:positronfluxratios}.  Since the detection probability of neutrinos increases as $E_\nu^2$, this preferentially leads to detection of higher energy neutrinos. This detection bias reduces the fraction of positrons created in the detector from neutrinos originating at the cosmic noon relative to positrons from lower redshift neutrinos. However, both the number of detected positrons and the fraction of positrons arising from cosmic noon neutrinos increases with higher values of the CC SN temperature parameter $T$.
In fig.~\ref{fig:positronspectrum}, we also indicate thresholds for detection of the DSNB at SK/HK with and without  Gd loading. We can clearly see from the figure that lowering the threshold by loading the detectors with Gd will significantly enhance sensitivity to positrons initiated by neutrinos from the cosmic noon.

\begin{table}
\begin{centering}
\setlength{\tabcolsep}{0.42em}
\begin{tabular}{|c|c|c||c|c|}
  \hline
    \toprule
& \multicolumn{2}{c||}{Energy} & \multicolumn{2}{c|}{Energy Range} \\ \hline
Temp &10 MeV & 16 MeV  & $>$ 10 MeV & $>$ 16 MeV  \\ \hline
  4 MeV & 0.08 (0.17) & 0.01 (0.04) & 0.03 (0.08)  & 0.01 (0.02)\\ 
  6 MeV & 0.23 (0.34)  & 0.07 (0.16)  & 0.09 (0.18) & 0.04 (0.09) \\
  8 MeV & 0.34 (0.45)  & 0.17 (0.28) & 0.16 (0.28) & 0.10 (0.19) \\
  Mixed & 0.16 (0.28)  & 0.09 (0.19)  & 0.09 (0.19) & 0.06 (0.14) \\
  \hline
\end{tabular}
\caption{Table of positron fractions arising from the cosmic noon $(1.5\lesssim z \lesssim 3.5)$ neutrinos at HK assuming the UV/IR CSFH for a given energy/energy range (columns). The rows correspond to different assumptions of the CC SNe temperature spectral parameter ($T$).  Numbers in brackets indicate the corresponding fractions for the H$\alpha$ CSFH. For the energy ranges of the right two columns, the upper limit of the range has been taken to be 25 MeV. Higher values of $T$ would lead to a greater fraction of positrons arising from neutrinos from the cosmic noon.  From an experimental point-of-view, measurement of the positron spectrum with a lower threshold will receive greater contributions from the cosmic noon neutrinos and thus be more sensitive to the differences between the UV/IR and  H$\alpha$ CSFHs.}
\label{tab:positronfluxratios}
\end{centering}
\end{table}

\subsection{Background noise and detection threshold at Hyper-Kamiokande}
In order to discriminate between CSFHs with different SFRDs at the cosmic noon, it is advantageous to detect neutrinos at low energies which have a greater contribution from high-redshift supernovae. However, there are two factors working against the detection of low-energy neutrinos. First, as already mentioned, the detection probability of neutrinos increases as $E_\nu^2$, which preferentially leads to detection of higher energy neutrinos. Second, various backgrounds in the detector become important at low energies, overwhelming any chance of detecting the DSNB at energies $\lesssim 10$~MeV. While HK can in principle detect neutrinos down to 3~MeV of energy~\cite{Abe:2018uyc},  in practice the detector has several sources of backgrounds in the energy range of interest, which make extraction of the DSNB signal difficult and restrict the range of energies over which the spectrum can be studied.
Here, we list the primary sources of background at HK for the DSNB search:
\begin{enumerate}
\item Spallation from cosmic muons
\item Invisible muon background
\item Atmospheric $\bar{\nu}_e$, $\nu_e$ charged-current (CC) background
\item Atmospheric neutrino neutral current (NC) background
\item Reactor anti-neutrinos
\end{enumerate}

Several of these sources of background can be significantly reduced by cuts on the data and by loading Gd in the detector to tag neutrons in genuine inverse $\beta$-decay reactions.

Spallation from cosmic muons can be reduced considerably by applying suitable spatial and temporal coincidence cuts to remove events associated with visible muons entering the detector.  Loading Gd in the detector to tag neutrons significantly reduces the spallation background by many orders of magnitude. However, spallation products which are relatively long-lived and produce neutrons are harder to remove.  In ref.~\cite{Abe:2018uyc}, it was argued that of all spallation products $^9$Li with a half-life of 0.18~s is the most dominant residual spallation background.  However, the relatively short lifetime of $^9$Li allows for it to be rejected with 99.5\% probability using coincidence cuts and this could possibly be improved further by optimizing the cuts.

Atmospheric muon neutrinos can produce muons in the detector via CC interactions, and these muons are typically below the Cherenkov threshold. The decay of these ``invisible'' muons produces electrons/positrons, which creates a background that can mimic the signal. The positron spectrum from this background is the well known Michel spectrum and can therefore be modelled very accurately, and is also distinct from the expected DSNB positron spectrum. Furthermore, with $0.1\%$ Gd in the detector, the background from invisible muons can be further reduced by a factor of 5~\cite{Beacom:2003nk}.

Atmospheric $\bar{\nu}_e$ CC interactions produce a source of background which is indistinguishable from DSNB signal on an event-by-event basis, but fortunately has a very different spectrum compared to that expected from the DSNB. Atmospheric ${\nu}_e$ CC interactions produce electrons which are indistinguishable from positrons (and hence signal events) without Gd tagging, but can be reduced by looking for accompanying neutrons.

In addition, atmospheric neutrinos can interact elastically via neutral current interactions and some of these interactions may mimic the signal. The NC background can be reduced by Cerenkov angle cuts, but has still been shown to be important below positron energies of 20 MeV~\cite{Bays:2011si, Bays:2012wty}.

Reactor $\bar{\nu}_e$s are another source of irreducible background below neutrino energies of $\sim$ 10 MeV. While the background due to reactor neutrinos is overwhelming below 10 MeV, there is also a moderate tail from 10 - 15 MeV.

There are several other sources of background such as solar neutrinos and pions faking electrons, but these can be reduced to very low levels with solar angle and Cerenkov angle cuts respectively.

Thus, after application of the cuts and assuming neutron tagging with Gd loading at HK, the dominant surviving backgrounds in the energy range of interest are the $^9$Li spallation background, the unresolved reactor neutrino tail, invisible muon spallation, atmospheric $\nu_e/\bar{\nu}_e$ CC events and atmospheric NC events that survive the cuts. The spectra of these individual backgrounds (combined for both detectors) and their sum as a function of reconstructed energy is shown in fig.~\ref{fig:bgkds}.

We have taken our $^9$Li~\footnote{We have reduced the $^9$Li background shown in fig.~190 of ref.~\cite{Abe:2018uyc} by a factor of 4/6 based on a private correspondence with Hiroyuki Sekiya and Takatomi Yano.} and reactor neutrino backgrounds (assuming full reactor intensity) from the HK design report~\cite{Abe:2018uyc} and we have taken the NC background, the invisible muon background and the $\nu_e/\bar{\nu}_e$ CC backgrounds from an SK-Gd study~\cite{Sekiya:2020pun} appropriately scaled up to the HK detector volume.

The spallation background decreases with increased overburden, and with a 1 km depth for the proposed Korean sites at Mt Bisul and Mt Bohyun, the spallation background is expected to be a factor of $\sim$~4 smaller than at the Tochibora site~\cite{Abe:2016ero}. Since there is no study yet of the reactor neutrino background at the Korean sites, we take a nominal assumption of the same reactor background at both the Tochibora and Korean sites. Other backgrounds, sourced by atmospheric neutrinos, are expected to be the same at both sites.

At Super-Kamiokande without Gd loading, the spallation background limited the search for DSNB down to 17.3 MeV neutrino energy~\cite{Bays:2011si}.
The reduction of backgrounds with Gd loading allows for a lowering of the threshold energy of DSNB detection down to $\sim 10$~MeV, which is ultimately constrained by the reactor neutrino background. This reduction of the threshold will be extremely useful for studying neutrinos from the cosmic noon.

The cuts applied to the data to reduce the background also reduce the signal efficiency. While a detailed study of the reduced signal efficiency at a Gd loaded detector is not yet available, we assume a net efficiency of  $\epsilon_{\textrm{\tiny{eff}}}=70\%$, which is consistent with an efficiency of 78\% at SK~\cite{Bays:2011si} with a further 90\% reduction due to the neutron tagging efficiency.

\subsection{Analysis and Results}

\begin{figure}
    \centering
    \includegraphics[scale = 0.5]{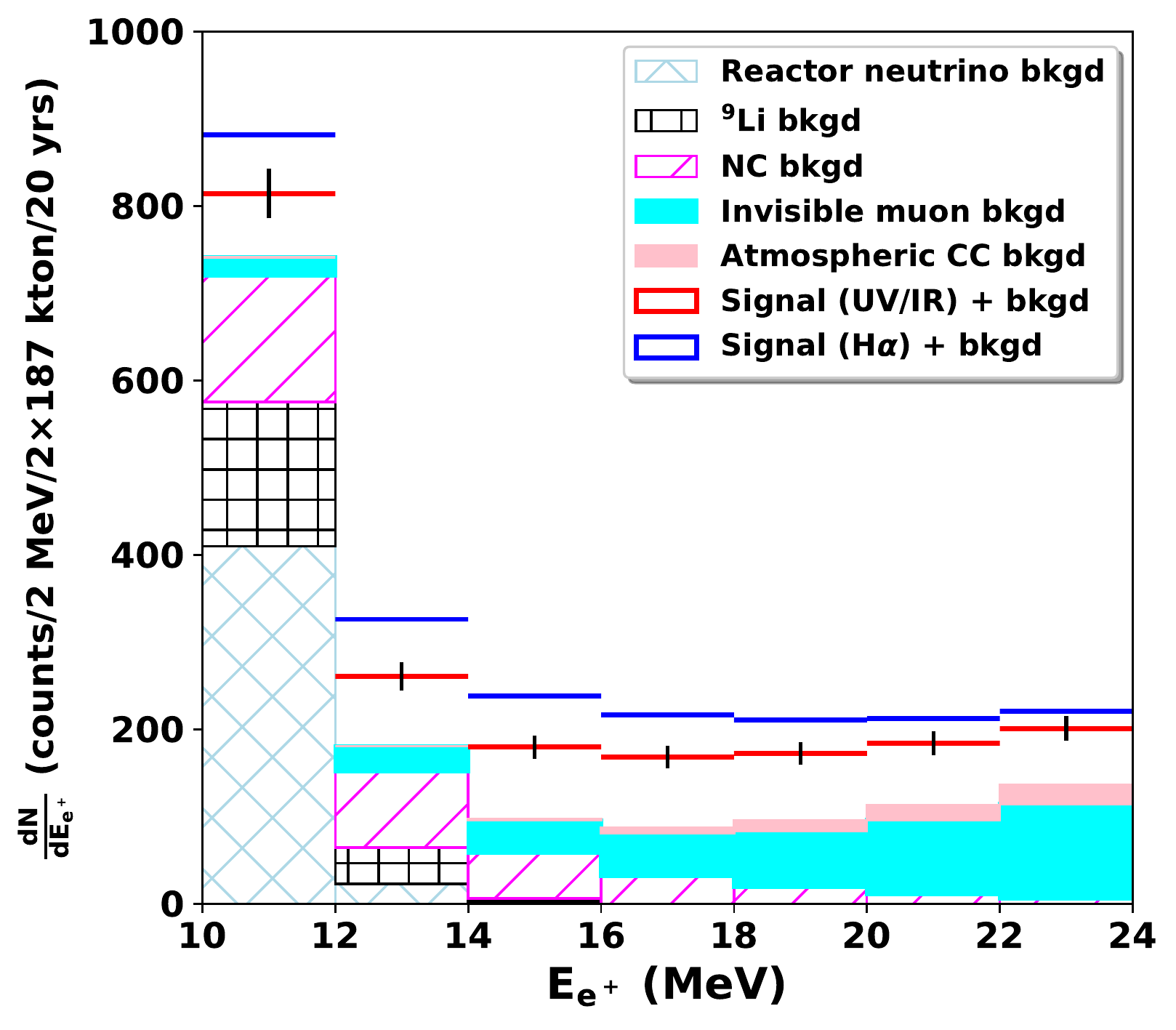}
    \caption{This histogram shows the expected backgrounds as a function of detected energy, combined from both HK sites with 20 years of data and assuming Gd loading. The combined DSNB signal with background is shown for both the UV/IR CSFH and the H$\alpha$ CSFH assuming a CC SNe spectral parameter of $T = 8$~MeV. We have also shown the Poissonian ($\sqrt{N}$) errors for the UV/IR CSFH case for reference. We set 10~MeV as the lower threshold for our analysis to discriminate between the two different signal hypotheses, below which the reactor neutrino background overwhelms the signal.}
    \label{fig:bgkds}
\end{figure}

In fig.~\ref{fig:bgkds}, we show a histogram of the expected positron signal with background for the UV/IR and H$\alpha$ CSFHs assuming a spectral parameter $T=8$~MeV for the CC SNe temperature. The histogram represents expected data combined from both HK experiments over 20 years and also shows the relative contributions of different sources of background. We have also shown the Poissonian ($\sqrt{N}$) statistical error bars on the expected data assuming the UV/IR CSFH.

The difference between DSNB signal and background is prominent over a large energy range from $10 - 40$~MeV. However, the difference between the two signal hypothesis is appreciable only below about 24~MeV and is more prominent at lower energies, where the difference in cosmic noon neutrino contributions is largest.

Given an assumption of the true CSFH. We can now ask two questions: a) How many years of data collection would it take to discover the DSNB signal? b) How many years of data collection would it take to rule out the alternative hypothesis for the CSFH? For concreteness, we will take the specific case of the assumption that the UV/IR CSFH is the correct one below.

To answer both the questions above, we need to define a null and an alternative hypothesis and a test statistic that will discriminate between them. For discovery potential, we take the null hypothesis to be background only and the alternative hypothesis to be background with DSNB signal assuming the UV/IR CSFH. For excluding the H$\alpha$ CSFH, we take our null hypothesis to be background with DSNB signal assuming the H$\alpha$ CSFH and our alternative hypothesis to be  background with DSNB signal assuming the UV/IR CSFH.
We define a chi-squared test statistic in the usual way as,
\begin{equation}
\chi^2 \equiv \sum_{i} \left( \frac{\textrm{obs}_i - \textrm{pred}_i}{\sigma_i} \right )^2,
\end{equation}
where $\textrm{obs}_i$ is the observed (pseudo-)data in the $i$-th energy bin and $\textrm{pred}_i$ is the mean predicted number of events under the null hypothesis in the same energy bin and $\sigma_i = \sqrt{\textrm{pred}_i}$, is the Poissonian fluctuation in the predicted number of events. Here, we have taken 2 MeV bins starting from an energy threshold of 10 MeV going up to 24 MeV, where the lower threshold is set by the reactor neutrino background, and the upper threshold is chosen in order to select a range where the two CSFHs yield the most different positron spectra~\footnote{For discovery of the DSNB it will be useful to extend the energy range of analysis to higher energies, but for simplicity, since we are focussed on cosmic noon neutrinos, we will use the same energy range for both discovery and exclusion analysis.}.

We then calculate the \textit{average} chi-squared $\langle \chi^2 \rangle$ and its 1-$\sigma$ fluctuation $\Delta \chi^2\equiv\sqrt{\langle (\chi^2 - \langle \chi^2 \rangle)^2\rangle}$. We will denote the expected mean number of events in the $i$-th bin under the null and alternative hypothesis as $N_i$ and $A_i$, respectively. Then, fixing $\textrm{pred}_i =N_i$ and averaging over Poissonian, independent fluctuations of $\textrm{obs}_i$ about the mean value of $A_i$, we obtain the following analytic expressions for  $\langle \chi^2 \rangle$ and $\Delta \chi^2$,
\begin{gather}
\langle \chi^2 \rangle = \sum_{i} \left [\frac{ A_i}{N_i}  + \left (\frac{(A_i - N_i)^2}{N_i} \right ) \right ], \\
 \Delta \chi^2 = \sqrt{\sum_{i} \left[\frac{4 A_i [ (A_i - N_i)^2 + (A_i - N_i)] + 2 A_i^2 + A_i}{N_i^2}    \right ]}.
\end{gather}
Note that both  $N_i$ and $A_i$ scale linearly with the number of years of data collection.
We can now convert $\langle \chi^2 \rangle$ and its fluctuations $\langle \chi^2 \rangle  \pm \Delta \chi^2$ into an average $p$-value $\langle p\rangle$ for discovery/exclusion and its fluctuation $\langle p   \rangle \pm \Delta p$, for a given number of years of data collection.

We can then proceed to answer our questions above. The expected number of years required to discover the DSNB, assuming the UV/IR CSFH, would be the number of years needed for the average $p$-value to yield a 5-$\sigma$ exclusion of the null hypothesis of background only, i.e. $\langle p   \rangle =6\times 10^{-7}$. We can also find the fluctuation on the expected number of years for discovery by requiring $\langle p  \rangle \pm \Delta p  = 6\times 10^{-7}$. Similarly, we can estimate the time required to rule out the alternative H$\alpha$ CSFH at the 2-$\sigma$ level ($p=0.05$), and its uncertainty.

In table~\ref{tab:UVIRdiscoveryexclusion}, we show the expected number of years to discovery of the DSNB signal and exclusion of the H$\alpha$ CSFH  assuming that the UV/IR CSFH is the right one under different assumptions of the CC SNe temperature parameter.

\begin{table}
\begin{centering}
\setlength{\tabcolsep}{0.42em}
\def\arraystretch{1.4}%
\begin{tabular}{|c||c|c|}
  \hline
    \toprule
Temp & {5-$\sigma$ Discovery} &{2-$\sigma$  Exclusion of H$\alpha$ CSFH} \\  \hhline{|=#=|=|}
4 MeV &10.7$^{+5.6}_{-3.9}$ & 123.4$^{+154.0}_{-84.5}$  \\ \hline
6 MeV &2.9$^{+1.8}_{-1.2}$ &  8.2$^{+8.6}_{-4.9}$  \\ \hline
8 MeV &1.9$^{+1.3}_{-0.8}$ & 2.9$^{+2.7}_{-1.6}$   \\ \hline
Mixed &5.6$^{+3.2}_{-2.2}$ & 19.4$^{+21.3}_{-11.9}$  \\ \hline
\end{tabular}
\caption{Left column: Years needed for 5-$\sigma$ discovery of the DSNB at HK loaded with Gd assuming that the UV/IR inferred CSFH is correct. Right column: Years needed for 2-$\sigma$ exclusion of the H$\alpha$ CSFH assuming that the UV/IR inferred CSFH is correct. Rows correspond to different assumptions of the CC SNe temperature spectral parameter.}
\label{tab:UVIRdiscoveryexclusion}
\end{centering}
\end{table}

In table~\ref{tab:Halphadiscoveryexclusion}, we show the expected number of years to discovery of the DSNB signal and exclusion of the UV/IR CSFH  assuming that the H$\alpha$ CSFH is the right one under different assumptions of the CC SNe temperature parameter.

\begin{table}
\begin{centering}
\setlength{\tabcolsep}{0.42em}
\def\arraystretch{1.4}%
\begin{tabular}{|c||c|c|}
  \hline
    \toprule
Temp & {5-$\sigma$ Discovery} &{2-$\sigma$  Exclusion of UV/IR CSFH} \\ \hhline{|=#=|=|}
4 MeV &9.8$^{+5.1}_{-3.6}$ & 123.4$^{+153.7}_{-83.4}$   \\ \hline
6 MeV &1.5$^{+1.1}_{-0.7}$ & 6.0$^{+9.1}_{-4.8}$ \\ \hline
8 MeV &0.7$^{+0.6}_{-0.4}$ & 1.6$^{+3.0}_{-1.6}$  \\ \hline
Mixed &3.2$^{+1.9}_{-1.3}$ & 15.5$^{+22.1}_{-11.9}$  \\
  \hline
\end{tabular}
\caption{Left column: Years needed for 5-$\sigma$ discovery of the DSNB at HK loaded with Gd assuming that the H$\alpha$ inferred CSFH is correct. Right column: Years needed for 2-$\sigma$ exclusion of the UV/IR CSFH assuming that the H$\alpha$ inferred CSFH is correct. Rows correspond to different assumptions of the CC SNe temperature spectral parameter.}
\label{tab:Halphadiscoveryexclusion}
\end{centering}
\end{table}

From the tables, we can see that for either the UV/IR or H$\alpha$ assumption of the CSFH,  discovery of the DSNB at HK loaded with Gd should take less than 10 years. Even with a low CC SNe temperature parameter of 4~MeV, the DSNB should be discoverable at HK, but the lower threshold sensitivity due to the addition of Gd will be critical for discovery with this relatively lower energy spectrum. Unsurprisingly, the discovery of the DSNB, assuming the H$\alpha$ CSFH is correct, is expected to happen sooner given the larger signal rates.

Now we come to the central question raised in the introduction: Can the DSNB be used to probe the difference between the CSFH inferred from UV/IR data and H$\alpha$ data? We find that the DSNB signal at HK could potentially exclude the UV/IR or H$\alpha$ CSFH hypothesis in favor of the other within 1.6-2.9 years of data collection in the scenario where the CC SNe temperature parameter $T=8$ MeV. In fact, for all our benchmark temperature hypotheses (except the $T= 4$~MeV scenario), we find that one of the CSFHs can be ruled out within 20 years of data collection. For the case where $T= 4$~MeV, the DSNB signals above the 10 MeV threshold are almost identical and the signal is relatively weak, thus discovery is possible, although discrimination of the different CSFHs will not be possible for any reasonable time-line. However, for even a small (30\%) admixture of $T=8$~MeV CC SNe, it will be possible to discriminate between the UV/IR and H$\alpha$ CSFH within 20 years of data collection.

Thus, the DNSB is a powerful tool to probe the CSFH and its potential detection and characterization at HK will be invaluable to discriminate between the different CSFHs inferred from UV/IR data and H$\alpha$ data. A critical requirement for this discrimination is the low threshold of detection in order to be sensitive to a larger fraction of neutrinos from the cosmic noon. This low threshold is possible at HK detectors loaded with Gd to reduce backgrounds.

While our analysis has focussed on HK loaded with Gd, we have also performed similar analyses for SK-Gd and HK without gadolinium (and assuming no neutron tagging, for e.g. by considering neutron capture on hydrogen) taking into account appropriate backgrounds for each detector. We attempted to understand whether the discovery of the DSNB and the rejection of the alternative CSFH can be made over a 20-year time span.

For SK-Gd, we have included reactor neutrino backgrounds and $^9$Li backgrounds in addition to the backgrounds considered in ref.~\cite{Sekiya:2020pun} and we find that discovery of the DSNB is only possible for the benchmark scenario where $T=8$~MeV and also if the true CSFH is the one inferred from H$\alpha$ data. We find in this case that 5-$\sigma$ discovery is possible within $11.0^{+9.4}_{-6.0}$ years, but exclusion of the UV/IR CSFH is not possible over a 20-year time span.

For HK without Gd, we have considered an analysis threshold of 16~MeV to reject the increased spallation background that would be present. The higher threshold would imply a lower sensitivity to cosmic noon neutrinos. We find that discovery is possible for $T=6$~MeV with a mean time of 17.5$^{+8.8}_{-6.0}$ (11.8$^{+6.1}_{-4.2}$) years for the UV/IR (H$\alpha$) CSFH, and also for $T=8$~MeV with a mean time of 9.8$^{+5.1}_{-3.5}$ (4.3$^{+2.4}_{-1.6}$) years.
However, exclusion of the alternative CSFH is only possible for the $T=8$~MeV scenario, and we find the mean time to 2-$\sigma$ exclusion to be 10.5$^{+13.3}_{-6.7}$ (8.5$^{+13.8}_{-6.6}$) years assuming the underlying CSFH is the UV/IR (H$\alpha$) one.

Thus, at SK-Gd discovery of the DSNB will be possible only with a fortuitous combination of CSFH and temperature of the CC SNe, and in such a situation HK without Gd has the potential to study the spectrum and discriminate between the UV/IR and H$\alpha$ CSFH. However, for the broader set of possible CC SNe spectra, it is crucial that HK is loaded with Gd to allow a study of the spectrum of CC SNe down to $\sim$ 10 MeV, which would permit an exploration of cosmic noon neutrinos and discriminate between the UV/IR and H$\alpha$ CSFHs. \textit{Thus, both the larger detector volume of HK and the lowered threshold with Gd loading are required to exclude either the H$\alpha$ or UV/IR CSFH.}

\section{Summary and conclusions}
Understanding galaxy formation and evolution is one of the open frontier problems in modern cosmology and is a critical step towards a better understanding of the behaviour of dark matter on small scales. One of the key observational inputs needed to study galaxy formation and evolution is knowledge of the cosmic star formation history.

Observations using multiple probes of the CSFH in different wavelength bands of the electromagnetic spectrum indicate an epoch of peak star formation between $1.5 \lesssim z \lesssim 3.5$ known as the \textit{cosmic noon}. These probes approximately agree on the redshift of the cosmic noon, but disagree on the peak magnitude of star formation. In particular,  H$\alpha$ data seems to indicate a peak SFRD which is almost four times larger than the peak indicated by UV/IR data. It is possible that this discrepancy could be resolved by calibration uncertainties and redshift dependence of the conversion factors used to map luminosity to SFRD. Underestimated UV extinction factors near cosmic noon and an underestimated AGN contribution to H$\alpha$ at high redshifts might reduce the discrepancy between the peak SFRDs. However, given the standard calibrations, it is interesting to ask if additional probes could be sensitive to the difference in the CSFH inferred from these probes.

We argued that the diffuse supernova neutrino background (DSNB) is a promising probe of the CSFH. In this work, the central question that we set out to address was whether a future detection and characterization of the DSNB spectrum could discriminate between the CSFHs inferred from UV/IR data and H$\alpha$ data.

Since the spectrum of cosmologically averaged CC SNe is not yet known, for our analysis, we selected a set of benchmark CC~SNe spectra which differed only in their choice of temperature parameter ($T$), with higher temperatures leading to a slightly higher energy spectrum. Numerical simulations, possibly calibrated to future observations of local SNe (within 10 Mpc), could solidify our understanding of this spectrum within the next decade or two.

Since cosmological neutrinos are redshifted, the low energy DSNB is more sensitive to differences in the SFRD from the cosmic noon. However, there are several challenges with detection of these cosmic noon neutrinos:
\begin{itemize}
\item The detection probability of neutrinos scales as $E_\nu^2$, which suppresses the rate of lower energy neutrino signals relative to higher energy neutrinos.
\item Various detector backgrounds make the study of low energy neutrinos challenging.
\end{itemize}

We then studied the potential of the next-generation Hyper-Kamiokande detector to characterize the spectrum of the DSNB and to discriminate between different CSFHs. The baseline design for HK only looks for positrons produced in inverse $\beta$-decay reactions and will be limited in its low energy threshold to $\sim$ 16 MeV due to spallation backgrounds. However, loading HK with 0.1\% gadolinium by mass to tag accompanying neutrons can lower the spallation background (and other backgrounds) significantly, lowering the threshold to  $\sim$ 10 MeV, below which reactor neutrino background overwhelms any signal.

We found that depending on the CC SNe spectral parameters, discovery of the DSNB is almost assured over a 10-year time scale, and possibly even within the first year of operation of HK loaded with gadolinium. Our main result was to show that discrimination of the CSFH inferred from UV/IR data versus that inferred from H$\alpha$ data should be possible within 1.6-20 years of operation depending on the $T$ parameter. However, in the benchmark scenario where $T = 4$~MeV, we find that although the DSNB should be discoverable,  the spectrum will be too low in energy for HK to be sensitive to the cosmic noon neutrinos and it will not be able to discriminate between the different CSFHs on a reasonable timescale.

Without Gd loading at HK, we find that discrimination of the two CSFHs, over a 10-20 year time scale, is only possible in the most optimistic benchmark scenario where $T=8$~MeV, although discovery of the DSNB signal may be possible with $T=6$~MeV over this timescale.

We have also found that lower volume detectors such as SK-Gd, even with a low threshold sensitivity, will not have sufficient statistics to discriminate between the two CSFHs.

Our results have only taken into account statistical uncertainties and we have ignored systematic uncertainties which might affect our conclusions. Some of these important systematic uncertainties would be the uncertainty on various low energy backgrounds, and crucially, uncertainties on the to-be-determined CC SNe spectral parameters.

Calibration of numerical simulations of CC~SNe can be accomplished using galactic SNe at current detectors (assuming we are lucky enough to see an explosion) or to observations of nearby SNe, which would require large volume detectors like HK itself. In the latter case, characterizing the spectrum using HK data would require analysis and extraction of local SNe spectral information, only after which, one can use the simultaneously collected DSNB data to discriminate between the two CSFHs. Thus, unless we are fortunate enough to see a galactic SN explosion soon, discrimination of the two CSFHs will require waiting until sufficient data is gathered to understand the CC~SN spectrum reliably.

In conclusion, the ability of a next-generation, large-volume detector like HK to resolve a discrepancy in the CSFH based on conventional electromagnetic probes would be a significant milestone for neutrino astrophysics. In order to be sensitive to the cosmic noon neutrinos for the broadest range of CC SNe spectra, there is a clear need for Hyper-Kamiokande to go to lower thresholds and this can be accomplished via gadolinium loading. We hope that future detector commissioning plans for HK will take into account this major science goal.

\section*{Acknowledgements}
The authors acknowledge very helpful correspondence with Hiroyuki Sekiya and Takatomi Yano regarding backgrounds at HK. The authors thank Varun Bhalerao and Basudeb Dasgupta for enlightening discussions and helpful comments on the draft. VR is supported by a DST-SERB Early Career Research Award (ECR/2017/000040) and an IITB-IRCC seed grant.

\bibliography{SFRDSNB}

\begin{thebibliography}{95}
\expandafter\ifx\csname natexlab\endcsname\relax\def\natexlab#1{#1}\fi
\expandafter\ifx\csname bibnamefont\endcsname\relax
  \def\bibnamefont#1{#1}\fi
\expandafter\ifx\csname bibfnamefont\endcsname\relax
  \def\bibfnamefont#1{#1}\fi
\expandafter\ifx\csname citenamefont\endcsname\relax
  \def\citenamefont#1{#1}\fi
\expandafter\ifx\csname url\endcsname\relax
  \def\url#1{\texttt{#1}}\fi
\expandafter\ifx\csname urlprefix\endcsname\relax\def\urlprefix{URL }\fi
\providecommand{\bibinfo}[2]{#2}
\providecommand{\eprint}[2][]{\url{#2}}

\bibitem[{\citenamefont{Aghanim et~al.}(2018)}]{Aghanim:2018eyx}
\bibinfo{author}{\bibfnamefont{N.}~\bibnamefont{Aghanim}} \bibnamefont{et~al.}
  (\bibinfo{collaboration}{Planck}) (\bibinfo{year}{2018}),
  \eprint{1807.06209}.

\bibitem[{\citenamefont{Percival}(2015)}]{Percival:2013awa}
\bibinfo{author}{\bibfnamefont{W.~J.} \bibnamefont{Percival}}, in
  \emph{\bibinfo{booktitle}{{Proceedings, International School of Physics
  'Enrico Fermi': New Horizons for Observational Cosmology: Rome, Italy, June
  30-July 6, 2013}}} (\bibinfo{year}{2015}), vol. \bibinfo{volume}{186}, pp.
  \bibinfo{pages}{101--135}, \bibinfo{note}{[,317(2015)]}, \eprint{1312.5490}.

\bibitem[{\citenamefont{Silk}(2017)}]{Silk:2016srn}
\bibinfo{author}{\bibfnamefont{J.}~\bibnamefont{Silk}}, \bibinfo{journal}{JPS
  Conf. Proc.} \textbf{\bibinfo{volume}{14}}, \bibinfo{pages}{010101}
  (\bibinfo{year}{2017}), \eprint{1611.09846}.

\bibitem[{\citenamefont{Vogelsberger et~al.}(2019)\citenamefont{Vogelsberger,
  Marinacci, Torrey, and Puchwein}}]{Vogelsberger:2019ynw}
\bibinfo{author}{\bibfnamefont{M.}~\bibnamefont{Vogelsberger}},
  \bibinfo{author}{\bibfnamefont{F.}~\bibnamefont{Marinacci}},
  \bibinfo{author}{\bibfnamefont{P.}~\bibnamefont{Torrey}}, \bibnamefont{and}
  \bibinfo{author}{\bibfnamefont{E.}~\bibnamefont{Puchwein}}
  (\bibinfo{year}{2019}), \eprint{1909.07976}.

\bibitem[{\citenamefont{Oh et~al.}(2011)\citenamefont{Oh, de~Blok, Brinks,
  Walter, and Kennicutt}}]{Oh:2010ea}
\bibinfo{author}{\bibfnamefont{S.-H.} \bibnamefont{Oh}},
  \bibinfo{author}{\bibfnamefont{W.~J.~G.} \bibnamefont{de~Blok}},
  \bibinfo{author}{\bibfnamefont{E.}~\bibnamefont{Brinks}},
  \bibinfo{author}{\bibfnamefont{F.}~\bibnamefont{Walter}}, \bibnamefont{and}
  \bibinfo{author}{\bibfnamefont{R.~C.} \bibnamefont{Kennicutt},
  \bibfnamefont{Jr}}, \bibinfo{journal}{Astron. J.}
  \textbf{\bibinfo{volume}{141}}, \bibinfo{pages}{193} (\bibinfo{year}{2011}),
  \eprint{1011.0899}.

\bibitem[{\citenamefont{Rocha et~al.}(2013)\citenamefont{Rocha, Peter, Bullock,
  Kaplinghat, Garrison-Kimmel, Onorbe, and Moustakas}}]{Rocha:2012jg}
\bibinfo{author}{\bibfnamefont{M.}~\bibnamefont{Rocha}},
  \bibinfo{author}{\bibfnamefont{A.~H.~G.} \bibnamefont{Peter}},
  \bibinfo{author}{\bibfnamefont{J.~S.} \bibnamefont{Bullock}},
  \bibinfo{author}{\bibfnamefont{M.}~\bibnamefont{Kaplinghat}},
  \bibinfo{author}{\bibfnamefont{S.}~\bibnamefont{Garrison-Kimmel}},
  \bibinfo{author}{\bibfnamefont{J.}~\bibnamefont{Onorbe}}, \bibnamefont{and}
  \bibinfo{author}{\bibfnamefont{L.~A.} \bibnamefont{Moustakas}},
  \bibinfo{journal}{Mon. Not. Roy. Astron. Soc.}
  \textbf{\bibinfo{volume}{430}}, \bibinfo{pages}{81} (\bibinfo{year}{2013}),
  \eprint{1208.3025}.

\bibitem[{\citenamefont{Peter et~al.}(2013)\citenamefont{Peter, Rocha, Bullock,
  and Kaplinghat}}]{Peter:2012jh}
\bibinfo{author}{\bibfnamefont{A.~H.~G.} \bibnamefont{Peter}},
  \bibinfo{author}{\bibfnamefont{M.}~\bibnamefont{Rocha}},
  \bibinfo{author}{\bibfnamefont{J.~S.} \bibnamefont{Bullock}},
  \bibnamefont{and}
  \bibinfo{author}{\bibfnamefont{M.}~\bibnamefont{Kaplinghat}},
  \bibinfo{journal}{Mon. Not. Roy. Astron. Soc.}
  \textbf{\bibinfo{volume}{430}}, \bibinfo{pages}{105} (\bibinfo{year}{2013}),
  \eprint{1208.3026}.

\bibitem[{\citenamefont{{Zavala} et~al.}(2013)\citenamefont{{Zavala},
  {Vogelsberger}, and {Walker}}}]{2013MNRAS.431L..20Z}
\bibinfo{author}{\bibfnamefont{J.}~\bibnamefont{{Zavala}}},
  \bibinfo{author}{\bibfnamefont{M.}~\bibnamefont{{Vogelsberger}}},
  \bibnamefont{and} \bibinfo{author}{\bibfnamefont{M.~G.}
  \bibnamefont{{Walker}}}, \bibinfo{journal}{\mnras}
  \textbf{\bibinfo{volume}{431}}, \bibinfo{pages}{L20} (\bibinfo{year}{2013}),
  \eprint{1211.6426}.

\bibitem[{\citenamefont{Boylan-Kolchin
  et~al.}(2011)\citenamefont{Boylan-Kolchin, Bullock, and
  Kaplinghat}}]{BoylanKolchin:2011de}
\bibinfo{author}{\bibfnamefont{M.}~\bibnamefont{Boylan-Kolchin}},
  \bibinfo{author}{\bibfnamefont{J.~S.} \bibnamefont{Bullock}},
  \bibnamefont{and}
  \bibinfo{author}{\bibfnamefont{M.}~\bibnamefont{Kaplinghat}},
  \bibinfo{journal}{Mon.Not.Roy.Astron.Soc.} \textbf{\bibinfo{volume}{415}},
  \bibinfo{pages}{L40} (\bibinfo{year}{2011}), \eprint{1103.0007}.

\bibitem[{\citenamefont{{Vogelsberger}
  et~al.}(2012)\citenamefont{{Vogelsberger}, {Zavala}, and
  {Loeb}}}]{2012MNRAS.423.3740V}
\bibinfo{author}{\bibfnamefont{M.}~\bibnamefont{{Vogelsberger}}},
  \bibinfo{author}{\bibfnamefont{J.}~\bibnamefont{{Zavala}}}, \bibnamefont{and}
  \bibinfo{author}{\bibfnamefont{A.}~\bibnamefont{{Loeb}}},
  \bibinfo{journal}{\mnras} \textbf{\bibinfo{volume}{423}},
  \bibinfo{pages}{3740} (\bibinfo{year}{2012}), \eprint{1201.5892}.

\bibitem[{\citenamefont{Bullock et~al.}(2000)\citenamefont{Bullock, Kravtsov,
  and Weinberg}}]{Bullock:2000wn}
\bibinfo{author}{\bibfnamefont{J.~S.} \bibnamefont{Bullock}},
  \bibinfo{author}{\bibfnamefont{A.~V.} \bibnamefont{Kravtsov}},
  \bibnamefont{and} \bibinfo{author}{\bibfnamefont{D.~H.}
  \bibnamefont{Weinberg}}, \bibinfo{journal}{Astrophys. J.}
  \textbf{\bibinfo{volume}{539}}, \bibinfo{pages}{517} (\bibinfo{year}{2000}),
  \eprint{astro-ph/0002214}.

\bibitem[{\citenamefont{Benson et~al.}(2002)\citenamefont{Benson, Frenk, Lacey,
  Baugh, and Cole}}]{Benson:2001at}
\bibinfo{author}{\bibfnamefont{A.~J.} \bibnamefont{Benson}},
  \bibinfo{author}{\bibfnamefont{C.~S.} \bibnamefont{Frenk}},
  \bibinfo{author}{\bibfnamefont{C.~G.} \bibnamefont{Lacey}},
  \bibinfo{author}{\bibfnamefont{C.~M.} \bibnamefont{Baugh}}, \bibnamefont{and}
  \bibinfo{author}{\bibfnamefont{S.}~\bibnamefont{Cole}},
  \bibinfo{journal}{Mon. Not. Roy. Astron. Soc.}
  \textbf{\bibinfo{volume}{333}}, \bibinfo{pages}{177} (\bibinfo{year}{2002}),
  \eprint{astro-ph/0108218}.

\bibitem[{\citenamefont{{Governato} et~al.}(2012)\citenamefont{{Governato},
  {Zolotov}, {Pontzen}, {Christensen}, {Oh}, {Brooks}, {Quinn}, {Shen}, and
  {Wadsley}}}]{2012MNRAS.422.1231G}
\bibinfo{author}{\bibfnamefont{F.}~\bibnamefont{{Governato}}},
  \bibinfo{author}{\bibfnamefont{A.}~\bibnamefont{{Zolotov}}},
  \bibinfo{author}{\bibfnamefont{A.}~\bibnamefont{{Pontzen}}},
  \bibinfo{author}{\bibfnamefont{C.}~\bibnamefont{{Christensen}}},
  \bibinfo{author}{\bibfnamefont{S.~H.} \bibnamefont{{Oh}}},
  \bibinfo{author}{\bibfnamefont{A.~M.} \bibnamefont{{Brooks}}},
  \bibinfo{author}{\bibfnamefont{T.}~\bibnamefont{{Quinn}}},
  \bibinfo{author}{\bibfnamefont{S.}~\bibnamefont{{Shen}}}, \bibnamefont{and}
  \bibinfo{author}{\bibfnamefont{J.}~\bibnamefont{{Wadsley}}},
  \bibinfo{journal}{\mnras} \textbf{\bibinfo{volume}{422}},
  \bibinfo{pages}{1231} (\bibinfo{year}{2012}), \eprint{1202.0554}.

\bibitem[{\citenamefont{Del~Popolo and Le~Delliou}(2017)}]{DelPopolo:2016emo}
\bibinfo{author}{\bibfnamefont{A.}~\bibnamefont{Del~Popolo}} \bibnamefont{and}
  \bibinfo{author}{\bibfnamefont{M.}~\bibnamefont{Le~Delliou}},
  \bibinfo{journal}{Galaxies} \textbf{\bibinfo{volume}{5}}, \bibinfo{pages}{17}
  (\bibinfo{year}{2017}), \eprint{1606.07790}.

\bibitem[{\citenamefont{Madau and Dickinson}(2014)}]{Madau:2014bja}
\bibinfo{author}{\bibfnamefont{P.}~\bibnamefont{Madau}} \bibnamefont{and}
  \bibinfo{author}{\bibfnamefont{M.}~\bibnamefont{Dickinson}},
  \bibinfo{journal}{Ann. Rev. Astron. Astrophys.}
  \textbf{\bibinfo{volume}{52}}, \bibinfo{pages}{415} (\bibinfo{year}{2014}),
  \eprint{1403.0007}.

\bibitem[{\citenamefont{Mac~Low}(2013)}]{MacLow1229229}
\bibinfo{author}{\bibfnamefont{M.-M.} \bibnamefont{Mac~Low}},
  \bibinfo{journal}{Science} \textbf{\bibinfo{volume}{340}}
  (\bibinfo{year}{2013}), ISSN \bibinfo{issn}{0036-8075}.

\bibitem[{\citenamefont{Hernquist and Springel}(2003)}]{Hernquist:2002rg}
\bibinfo{author}{\bibfnamefont{L.}~\bibnamefont{Hernquist}} \bibnamefont{and}
  \bibinfo{author}{\bibfnamefont{V.}~\bibnamefont{Springel}},
  \bibinfo{journal}{Mon. Not. Roy. Astron. Soc.}
  \textbf{\bibinfo{volume}{341}}, \bibinfo{pages}{1253} (\bibinfo{year}{2003}),
  \eprint{astro-ph/0209183}.

\bibitem[{\citenamefont{Schaye et~al.}(2015)}]{Schaye:2014tpa}
\bibinfo{author}{\bibfnamefont{J.}~\bibnamefont{Schaye}} \bibnamefont{et~al.},
  \bibinfo{journal}{Mon. Not. Roy. Astron. Soc.}
  \textbf{\bibinfo{volume}{446}}, \bibinfo{pages}{521} (\bibinfo{year}{2015}),
  \eprint{1407.7040}.

\bibitem[{\citenamefont{Pillepich et~al.}(2018)}]{Pillepich:2017jle}
\bibinfo{author}{\bibfnamefont{A.}~\bibnamefont{Pillepich}}
  \bibnamefont{et~al.}, \bibinfo{journal}{Mon. Not. Roy. Astron. Soc.}
  \textbf{\bibinfo{volume}{473}}, \bibinfo{pages}{4077} (\bibinfo{year}{2018}),
  \eprint{1703.02970}.

\bibitem[{\citenamefont{Conroy}(2013)}]{Conroy:2013if}
\bibinfo{author}{\bibfnamefont{C.}~\bibnamefont{Conroy}},
  \bibinfo{journal}{Ann. Rev. Astron. Astrophys.}
  \textbf{\bibinfo{volume}{51}}, \bibinfo{pages}{393} (\bibinfo{year}{2013}),
  \eprint{1301.7095}.

\bibitem[{\citenamefont{{Moustakas} et~al.}(2006)\citenamefont{{Moustakas},
  {Kennicutt}, and {Tremonti}}}]{2006ApJ...642..775M}
\bibinfo{author}{\bibfnamefont{J.}~\bibnamefont{{Moustakas}}},
  \bibinfo{author}{\bibfnamefont{J.}~\bibnamefont{{Kennicutt}},
  \bibfnamefont{Robert~C.}}, \bibnamefont{and}
  \bibinfo{author}{\bibfnamefont{C.~A.} \bibnamefont{{Tremonti}}},
  \bibinfo{journal}{\apj} \textbf{\bibinfo{volume}{642}}, \bibinfo{pages}{775}
  (\bibinfo{year}{2006}), \eprint{astro-ph/0511730}.

\bibitem[{\citenamefont{{Spergel} et~al.}(2015)\citenamefont{{Spergel},
  {Gehrels}, {Baltay}, {Bennett}, {Breckinridge}, {Donahue}, {Dressler},
  {Gaudi}, {Greene}, {Guyon} et~al.}}]{2015arXiv150303757S}
\bibinfo{author}{\bibfnamefont{D.}~\bibnamefont{{Spergel}}},
  \bibinfo{author}{\bibfnamefont{N.}~\bibnamefont{{Gehrels}}},
  \bibinfo{author}{\bibfnamefont{C.}~\bibnamefont{{Baltay}}},
  \bibinfo{author}{\bibfnamefont{D.}~\bibnamefont{{Bennett}}},
  \bibinfo{author}{\bibfnamefont{J.}~\bibnamefont{{Breckinridge}}},
  \bibinfo{author}{\bibfnamefont{M.}~\bibnamefont{{Donahue}}},
  \bibinfo{author}{\bibfnamefont{A.}~\bibnamefont{{Dressler}}},
  \bibinfo{author}{\bibfnamefont{B.~S.} \bibnamefont{{Gaudi}}},
  \bibinfo{author}{\bibfnamefont{T.}~\bibnamefont{{Greene}}},
  \bibinfo{author}{\bibfnamefont{O.}~\bibnamefont{{Guyon}}},
  \bibnamefont{et~al.}, \bibinfo{journal}{arXiv e-prints}
  \bibinfo{eid}{arXiv:1503.03757} (\bibinfo{year}{2015}), \eprint{1503.03757}.

\bibitem[{\citenamefont{{Green} et~al.}(2012)\citenamefont{{Green},
  {Schechter}, {Baltay}, {Bean}, {Bennett}, {Brown}, {Conselice}, {Donahue},
  {Fan}, {Gaudi} et~al.}}]{2012arXiv1208.4012G}
\bibinfo{author}{\bibfnamefont{J.}~\bibnamefont{{Green}}},
  \bibinfo{author}{\bibfnamefont{P.}~\bibnamefont{{Schechter}}},
  \bibinfo{author}{\bibfnamefont{C.}~\bibnamefont{{Baltay}}},
  \bibinfo{author}{\bibfnamefont{R.}~\bibnamefont{{Bean}}},
  \bibinfo{author}{\bibfnamefont{D.}~\bibnamefont{{Bennett}}},
  \bibinfo{author}{\bibfnamefont{R.}~\bibnamefont{{Brown}}},
  \bibinfo{author}{\bibfnamefont{C.}~\bibnamefont{{Conselice}}},
  \bibinfo{author}{\bibfnamefont{M.}~\bibnamefont{{Donahue}}},
  \bibinfo{author}{\bibfnamefont{X.}~\bibnamefont{{Fan}}},
  \bibinfo{author}{\bibfnamefont{B.~S.} \bibnamefont{{Gaudi}}},
  \bibnamefont{et~al.}, \bibinfo{journal}{arXiv e-prints}
  \bibinfo{eid}{arXiv:1208.4012} (\bibinfo{year}{2012}), \eprint{1208.4012}.

\bibitem[{\citenamefont{{Laureijs} et~al.}(2011)\citenamefont{{Laureijs},
  {Amiaux}, {Arduini}, {Augu{\`e}res}, {Brinchmann}, {Cole}, {Cropper},
  {Dabin}, {Duvet}, {Ealet} et~al.}}]{2011arXiv1110.3193L}
\bibinfo{author}{\bibfnamefont{R.}~\bibnamefont{{Laureijs}}},
  \bibinfo{author}{\bibfnamefont{J.}~\bibnamefont{{Amiaux}}},
  \bibinfo{author}{\bibfnamefont{S.}~\bibnamefont{{Arduini}}},
  \bibinfo{author}{\bibfnamefont{J.~L.} \bibnamefont{{Augu{\`e}res}}},
  \bibinfo{author}{\bibfnamefont{J.}~\bibnamefont{{Brinchmann}}},
  \bibinfo{author}{\bibfnamefont{R.}~\bibnamefont{{Cole}}},
  \bibinfo{author}{\bibfnamefont{M.}~\bibnamefont{{Cropper}}},
  \bibinfo{author}{\bibfnamefont{C.}~\bibnamefont{{Dabin}}},
  \bibinfo{author}{\bibfnamefont{L.}~\bibnamefont{{Duvet}}},
  \bibinfo{author}{\bibfnamefont{A.}~\bibnamefont{{Ealet}}},
  \bibnamefont{et~al.}, \bibinfo{journal}{arXiv e-prints}
  \bibinfo{eid}{arXiv:1110.3193} (\bibinfo{year}{2011}), \eprint{1110.3193}.

\bibitem[{\citenamefont{Pozzetti et~al.}(2016)\citenamefont{Pozzetti, Hirata,
  Geach, Cimatti, Baugh, Cucciati, Merson, Norberg, and
  Shi}}]{Pozzetti:2016cch}
\bibinfo{author}{\bibfnamefont{L.}~\bibnamefont{Pozzetti}},
  \bibinfo{author}{\bibfnamefont{C.~M.} \bibnamefont{Hirata}},
  \bibinfo{author}{\bibfnamefont{J.~E.} \bibnamefont{Geach}},
  \bibinfo{author}{\bibfnamefont{A.}~\bibnamefont{Cimatti}},
  \bibinfo{author}{\bibfnamefont{C.}~\bibnamefont{Baugh}},
  \bibinfo{author}{\bibfnamefont{O.}~\bibnamefont{Cucciati}},
  \bibinfo{author}{\bibfnamefont{A.}~\bibnamefont{Merson}},
  \bibinfo{author}{\bibfnamefont{P.}~\bibnamefont{Norberg}}, \bibnamefont{and}
  \bibinfo{author}{\bibfnamefont{D.}~\bibnamefont{Shi}},
  \bibinfo{journal}{Astron. Astrophys.} \textbf{\bibinfo{volume}{590}},
  \bibinfo{pages}{A3} (\bibinfo{year}{2016}), \eprint{1603.01453}.

\bibitem[{\citenamefont{Gardner et~al.}(2006)}]{Gardner:2006ky}
\bibinfo{author}{\bibfnamefont{J.~P.} \bibnamefont{Gardner}}
  \bibnamefont{et~al.}, \bibinfo{journal}{Space Sci. Rev.}
  \textbf{\bibinfo{volume}{123}}, \bibinfo{pages}{485} (\bibinfo{year}{2006}),
  \eprint{astro-ph/0606175}.

\bibitem[{\citenamefont{{Sobral} et~al.}(2013)\citenamefont{{Sobral}, {Smail},
  {Best}, {Geach}, {Matsuda}, {Stott}, {Cirasuolo}, and
  {Kurk}}}]{2013MNRAS.428.1128S}
\bibinfo{author}{\bibfnamefont{D.}~\bibnamefont{{Sobral}}},
  \bibinfo{author}{\bibfnamefont{I.}~\bibnamefont{{Smail}}},
  \bibinfo{author}{\bibfnamefont{P.~N.} \bibnamefont{{Best}}},
  \bibinfo{author}{\bibfnamefont{J.~E.} \bibnamefont{{Geach}}},
  \bibinfo{author}{\bibfnamefont{Y.}~\bibnamefont{{Matsuda}}},
  \bibinfo{author}{\bibfnamefont{J.~P.} \bibnamefont{{Stott}}},
  \bibinfo{author}{\bibfnamefont{M.}~\bibnamefont{{Cirasuolo}}},
  \bibnamefont{and} \bibinfo{author}{\bibfnamefont{J.}~\bibnamefont{{Kurk}}},
  \bibinfo{journal}{\mnras} \textbf{\bibinfo{volume}{428}},
  \bibinfo{pages}{1128} (\bibinfo{year}{2013}), \eprint{1202.3436}.

\bibitem[{\citenamefont{Gallego et~al.}(1995)\citenamefont{Gallego, Zamorano,
  Aragon-Salamanca, and Rego}}]{Gallego:1995ib}
\bibinfo{author}{\bibfnamefont{J.}~\bibnamefont{Gallego}},
  \bibinfo{author}{\bibfnamefont{J.}~\bibnamefont{Zamorano}},
  \bibinfo{author}{\bibfnamefont{A.}~\bibnamefont{Aragon-Salamanca}},
  \bibnamefont{and} \bibinfo{author}{\bibfnamefont{M.}~\bibnamefont{Rego}},
  \bibinfo{journal}{Astrophys. J.} \textbf{\bibinfo{volume}{455}},
  \bibinfo{pages}{L1} (\bibinfo{year}{1995}), \bibinfo{note}{[Astrophys.
  J.459,L43(1996)]}, \eprint{astro-ph/9510061}.

\bibitem[{\citenamefont{{Tresse} and {Maddox}}(1998)}]{1998ApJ...495..691T}
\bibinfo{author}{\bibfnamefont{L.}~\bibnamefont{{Tresse}}} \bibnamefont{and}
  \bibinfo{author}{\bibfnamefont{S.~J.} \bibnamefont{{Maddox}}},
  \bibinfo{journal}{\apj} \textbf{\bibinfo{volume}{495}}, \bibinfo{pages}{691}
  (\bibinfo{year}{1998}), \eprint{astro-ph/9709240}.

\bibitem[{\citenamefont{{Fujita} et~al.}(2003)\citenamefont{{Fujita}, {Ajiki},
  {Shioya}, {Nagao}, {Murayama}, {Taniguchi}, {Umeda}, {Yamada}, {Yagi},
  {Okamura} et~al.}}]{2003ApJ...586L.115F}
\bibinfo{author}{\bibfnamefont{S.~S.} \bibnamefont{{Fujita}}},
  \bibinfo{author}{\bibfnamefont{M.}~\bibnamefont{{Ajiki}}},
  \bibinfo{author}{\bibfnamefont{Y.}~\bibnamefont{{Shioya}}},
  \bibinfo{author}{\bibfnamefont{T.}~\bibnamefont{{Nagao}}},
  \bibinfo{author}{\bibfnamefont{T.}~\bibnamefont{{Murayama}}},
  \bibinfo{author}{\bibfnamefont{Y.}~\bibnamefont{{Taniguchi}}},
  \bibinfo{author}{\bibfnamefont{K.}~\bibnamefont{{Umeda}}},
  \bibinfo{author}{\bibfnamefont{S.}~\bibnamefont{{Yamada}}},
  \bibinfo{author}{\bibfnamefont{M.}~\bibnamefont{{Yagi}}},
  \bibinfo{author}{\bibfnamefont{S.}~\bibnamefont{{Okamura}}},
  \bibnamefont{et~al.}, \bibinfo{journal}{\apjl}
  \textbf{\bibinfo{volume}{586}}, \bibinfo{pages}{L115} (\bibinfo{year}{2003}),
  \eprint{astro-ph/0302473}.

\bibitem[{\citenamefont{{Hippelein} et~al.}(2003)\citenamefont{{Hippelein},
  {Maier}, {Meisenheimer}, {Wolf}, {Fried}, {von Kuhlmann}, {K{\"u}mmel},
  {Phleps}, and {R{\"o}ser}}}]{2003AnA...402...65H}
\bibinfo{author}{\bibfnamefont{H.}~\bibnamefont{{Hippelein}}},
  \bibinfo{author}{\bibfnamefont{C.}~\bibnamefont{{Maier}}},
  \bibinfo{author}{\bibfnamefont{K.}~\bibnamefont{{Meisenheimer}}},
  \bibinfo{author}{\bibfnamefont{C.}~\bibnamefont{{Wolf}}},
  \bibinfo{author}{\bibfnamefont{J.~W.} \bibnamefont{{Fried}}},
  \bibinfo{author}{\bibfnamefont{B.}~\bibnamefont{{von Kuhlmann}}},
  \bibinfo{author}{\bibfnamefont{M.}~\bibnamefont{{K{\"u}mmel}}},
  \bibinfo{author}{\bibfnamefont{S.}~\bibnamefont{{Phleps}}}, \bibnamefont{and}
  \bibinfo{author}{\bibfnamefont{H.~J.} \bibnamefont{{R{\"o}ser}}},
  \bibinfo{journal}{\aap} \textbf{\bibinfo{volume}{402}}, \bibinfo{pages}{65}
  (\bibinfo{year}{2003}), \eprint{astro-ph/0302116}.

\bibitem[{\citenamefont{{Hopkins}}(2004)}]{2004ApJ...615..209H}
\bibinfo{author}{\bibfnamefont{A.~M.} \bibnamefont{{Hopkins}}},
  \bibinfo{journal}{\apj} \textbf{\bibinfo{volume}{615}}, \bibinfo{pages}{209}
  (\bibinfo{year}{2004}), \eprint{astro-ph/0407170}.

\bibitem[{\citenamefont{{Hopkins} and {Beacom}}(2006)}]{2006ApJ...651..142H}
\bibinfo{author}{\bibfnamefont{A.~M.} \bibnamefont{{Hopkins}}}
  \bibnamefont{and} \bibinfo{author}{\bibfnamefont{J.~F.}
  \bibnamefont{{Beacom}}}, \bibinfo{journal}{\apj}
  \textbf{\bibinfo{volume}{651}}, \bibinfo{pages}{142} (\bibinfo{year}{2006}),
  \eprint{astro-ph/0601463}.

\bibitem[{\citenamefont{Shioya et~al.}(2008)}]{Shioya:2007kx}
\bibinfo{author}{\bibfnamefont{Y.}~\bibnamefont{Shioya}} \bibnamefont{et~al.},
  \bibinfo{journal}{Astrophys. J. Suppl.} \textbf{\bibinfo{volume}{175}},
  \bibinfo{pages}{128} (\bibinfo{year}{2008}), \eprint{0709.1009}.

\bibitem[{\citenamefont{{Ly} et~al.}(2007)\citenamefont{{Ly}, {Malkan},
  {Kashikawa}, {Shimasaku}, {Doi}, {Nagao}, {Iye}, {Kodama}, {Morokuma}, and
  {Motohara}}}]{2007ApJ...657..738L}
\bibinfo{author}{\bibfnamefont{C.}~\bibnamefont{{Ly}}},
  \bibinfo{author}{\bibfnamefont{M.~A.} \bibnamefont{{Malkan}}},
  \bibinfo{author}{\bibfnamefont{N.}~\bibnamefont{{Kashikawa}}},
  \bibinfo{author}{\bibfnamefont{K.}~\bibnamefont{{Shimasaku}}},
  \bibinfo{author}{\bibfnamefont{M.}~\bibnamefont{{Doi}}},
  \bibinfo{author}{\bibfnamefont{T.}~\bibnamefont{{Nagao}}},
  \bibinfo{author}{\bibfnamefont{M.}~\bibnamefont{{Iye}}},
  \bibinfo{author}{\bibfnamefont{T.}~\bibnamefont{{Kodama}}},
  \bibinfo{author}{\bibfnamefont{T.}~\bibnamefont{{Morokuma}}},
  \bibnamefont{and}
  \bibinfo{author}{\bibfnamefont{K.}~\bibnamefont{{Motohara}}},
  \bibinfo{journal}{\apj} \textbf{\bibinfo{volume}{657}}, \bibinfo{pages}{738}
  (\bibinfo{year}{2007}), \eprint{astro-ph/0610846}.

\bibitem[{\citenamefont{{Geach} et~al.}(2008)\citenamefont{{Geach}, {Smail},
  {Best}, {Kurk}, {Casali}, {Ivison}, and {Coppin}}}]{2008MNRAS.388.1473G}
\bibinfo{author}{\bibfnamefont{J.~E.} \bibnamefont{{Geach}}},
  \bibinfo{author}{\bibfnamefont{I.}~\bibnamefont{{Smail}}},
  \bibinfo{author}{\bibfnamefont{P.~N.} \bibnamefont{{Best}}},
  \bibinfo{author}{\bibfnamefont{J.}~\bibnamefont{{Kurk}}},
  \bibinfo{author}{\bibfnamefont{M.}~\bibnamefont{{Casali}}},
  \bibinfo{author}{\bibfnamefont{R.~J.} \bibnamefont{{Ivison}}},
  \bibnamefont{and} \bibinfo{author}{\bibfnamefont{K.}~\bibnamefont{{Coppin}}},
  \bibinfo{journal}{\mnras} \textbf{\bibinfo{volume}{388}},
  \bibinfo{pages}{1473} (\bibinfo{year}{2008}), \eprint{0805.2861}.

\bibitem[{\citenamefont{{Morioka} et~al.}(2008)\citenamefont{{Morioka},
  {Nakajima}, {Taniguchi}, {Shioya}, {Murayama}, and
  {Sasaki}}}]{2008PASJ...60.1219M}
\bibinfo{author}{\bibfnamefont{T.}~\bibnamefont{{Morioka}}},
  \bibinfo{author}{\bibfnamefont{A.}~\bibnamefont{{Nakajima}}},
  \bibinfo{author}{\bibfnamefont{Y.}~\bibnamefont{{Taniguchi}}},
  \bibinfo{author}{\bibfnamefont{Y.}~\bibnamefont{{Shioya}}},
  \bibinfo{author}{\bibfnamefont{T.}~\bibnamefont{{Murayama}}},
  \bibnamefont{and} \bibinfo{author}{\bibfnamefont{S.~S.}
  \bibnamefont{{Sasaki}}}, \bibinfo{journal}{\pasj}
  \textbf{\bibinfo{volume}{60}}, \bibinfo{pages}{1219} (\bibinfo{year}{2008}),
  \eprint{0807.0101}.

\bibitem[{\citenamefont{{Shim} et~al.}(2009)\citenamefont{{Shim}, {Colbert},
  {Teplitz}, {Henry}, {Malkan}, {McCarthy}, and {Yan}}}]{2009ApJ...696..785S}
\bibinfo{author}{\bibfnamefont{H.}~\bibnamefont{{Shim}}},
  \bibinfo{author}{\bibfnamefont{J.}~\bibnamefont{{Colbert}}},
  \bibinfo{author}{\bibfnamefont{H.}~\bibnamefont{{Teplitz}}},
  \bibinfo{author}{\bibfnamefont{A.}~\bibnamefont{{Henry}}},
  \bibinfo{author}{\bibfnamefont{M.}~\bibnamefont{{Malkan}}},
  \bibinfo{author}{\bibfnamefont{P.}~\bibnamefont{{McCarthy}}},
  \bibnamefont{and} \bibinfo{author}{\bibfnamefont{L.}~\bibnamefont{{Yan}}},
  \bibinfo{journal}{\apj} \textbf{\bibinfo{volume}{696}}, \bibinfo{pages}{785}
  (\bibinfo{year}{2009}), \eprint{0902.0736}.

\bibitem[{\citenamefont{Ly et~al.}(2010)\citenamefont{Ly, Lee, Dale, Momcheva,
  Salim, Staudaher, Moore, and Finn}}]{Ly_2010}
\bibinfo{author}{\bibfnamefont{C.}~\bibnamefont{Ly}},
  \bibinfo{author}{\bibfnamefont{J.~C.} \bibnamefont{Lee}},
  \bibinfo{author}{\bibfnamefont{D.~A.} \bibnamefont{Dale}},
  \bibinfo{author}{\bibfnamefont{I.}~\bibnamefont{Momcheva}},
  \bibinfo{author}{\bibfnamefont{S.}~\bibnamefont{Salim}},
  \bibinfo{author}{\bibfnamefont{S.}~\bibnamefont{Staudaher}},
  \bibinfo{author}{\bibfnamefont{C.~A.} \bibnamefont{Moore}}, \bibnamefont{and}
  \bibinfo{author}{\bibfnamefont{R.}~\bibnamefont{Finn}}, \bibinfo{journal}{The
  Astrophysical Journal} \textbf{\bibinfo{volume}{726}}, \bibinfo{pages}{109}
  (\bibinfo{year}{2010}),
  \urlprefix\url{https://doi.org/10.1088%2F0004-637x%2F726%2F2%2F109}.

\bibitem[{\citenamefont{{Tadaki} et~al.}(2011)\citenamefont{{Tadaki}, {Kodama},
  {Koyama}, {Hayashi}, {Tanaka}, and {Tokoku}}}]{2011PASJ...63S.437T}
\bibinfo{author}{\bibfnamefont{K.-I.} \bibnamefont{{Tadaki}}},
  \bibinfo{author}{\bibfnamefont{T.}~\bibnamefont{{Kodama}}},
  \bibinfo{author}{\bibfnamefont{Y.}~\bibnamefont{{Koyama}}},
  \bibinfo{author}{\bibfnamefont{M.}~\bibnamefont{{Hayashi}}},
  \bibinfo{author}{\bibfnamefont{I.}~\bibnamefont{{Tanaka}}}, \bibnamefont{and}
  \bibinfo{author}{\bibfnamefont{C.}~\bibnamefont{{Tokoku}}},
  \bibinfo{journal}{\pasj} \textbf{\bibinfo{volume}{63}}, \bibinfo{pages}{437}
  (\bibinfo{year}{2011}), \eprint{1012.4860}.

\bibitem[{\citenamefont{Gunawardhana et~al.}(2013)}]{Gunawardhana:2013fha}
\bibinfo{author}{\bibfnamefont{M.~L.~P.} \bibnamefont{Gunawardhana}}
  \bibnamefont{et~al.}, \bibinfo{journal}{Mon. Not. Roy. Astron. Soc.}
  \textbf{\bibinfo{volume}{433}}, \bibinfo{pages}{2764} (\bibinfo{year}{2013}),
  \eprint{1305.5308}.

\bibitem[{\citenamefont{{Stroe} and {Sobral}}(2015)}]{2015MNRAS.453..242S}
\bibinfo{author}{\bibfnamefont{A.}~\bibnamefont{{Stroe}}} \bibnamefont{and}
  \bibinfo{author}{\bibfnamefont{D.}~\bibnamefont{{Sobral}}},
  \bibinfo{journal}{\mnras} \textbf{\bibinfo{volume}{453}},
  \bibinfo{pages}{242} (\bibinfo{year}{2015}), \eprint{1507.02687}.

\bibitem[{\citenamefont{Sobral et~al.}(2015)}]{Sobral:2015tla}
\bibinfo{author}{\bibfnamefont{D.}~\bibnamefont{Sobral}} \bibnamefont{et~al.},
  \bibinfo{journal}{Mon. Not. Roy. Astron. Soc.}
  \textbf{\bibinfo{volume}{451}}, \bibinfo{pages}{2303} (\bibinfo{year}{2015}),
  \eprint{1502.06602}.

\bibitem[{\citenamefont{{Matthee} et~al.}(2017)\citenamefont{{Matthee},
  {Sobral}, {Best}, {Smail}, {Bian}, {Darvish}, {R{\"o}ttgering}, and
  {Fan}}}]{2017MNRAS.471..629M}
\bibinfo{author}{\bibfnamefont{J.}~\bibnamefont{{Matthee}}},
  \bibinfo{author}{\bibfnamefont{D.}~\bibnamefont{{Sobral}}},
  \bibinfo{author}{\bibfnamefont{P.}~\bibnamefont{{Best}}},
  \bibinfo{author}{\bibfnamefont{I.}~\bibnamefont{{Smail}}},
  \bibinfo{author}{\bibfnamefont{F.}~\bibnamefont{{Bian}}},
  \bibinfo{author}{\bibfnamefont{B.}~\bibnamefont{{Darvish}}},
  \bibinfo{author}{\bibfnamefont{H.}~\bibnamefont{{R{\"o}ttgering}}},
  \bibnamefont{and} \bibinfo{author}{\bibfnamefont{X.}~\bibnamefont{{Fan}}},
  \bibinfo{journal}{\mnras} \textbf{\bibinfo{volume}{471}},
  \bibinfo{pages}{629} (\bibinfo{year}{2017}), \eprint{1702.04721}.

\bibitem[{\citenamefont{{Hayashi} et~al.}(2018)\citenamefont{{Hayashi},
  {Tanaka}, {Shimakawa}, {Furusawa}, {Momose}, {Koyama}, {Silverman}, {Kodama},
  {Komiyama}, {Leauthaud} et~al.}}]{2018PASJ...70S..17H}
\bibinfo{author}{\bibfnamefont{M.}~\bibnamefont{{Hayashi}}},
  \bibinfo{author}{\bibfnamefont{M.}~\bibnamefont{{Tanaka}}},
  \bibinfo{author}{\bibfnamefont{R.}~\bibnamefont{{Shimakawa}}},
  \bibinfo{author}{\bibfnamefont{H.}~\bibnamefont{{Furusawa}}},
  \bibinfo{author}{\bibfnamefont{R.}~\bibnamefont{{Momose}}},
  \bibinfo{author}{\bibfnamefont{Y.}~\bibnamefont{{Koyama}}},
  \bibinfo{author}{\bibfnamefont{J.~D.} \bibnamefont{{Silverman}}},
  \bibinfo{author}{\bibfnamefont{T.}~\bibnamefont{{Kodama}}},
  \bibinfo{author}{\bibfnamefont{Y.}~\bibnamefont{{Komiyama}}},
  \bibinfo{author}{\bibfnamefont{A.}~\bibnamefont{{Leauthaud}}},
  \bibnamefont{et~al.}, \bibinfo{journal}{\pasj} \textbf{\bibinfo{volume}{70}},
  \bibinfo{eid}{S17} (\bibinfo{year}{2018}), \eprint{1704.05978}.

\bibitem[{\citenamefont{Coughlin et~al.}(2018)\citenamefont{Coughlin, Rhoads,
  Malhotra, Probst, Swaters, Tilvi, Zheng, Finkelstein, Hibon, Mobasher
  et~al.}}]{Coughlin_2018}
\bibinfo{author}{\bibfnamefont{A.}~\bibnamefont{Coughlin}},
  \bibinfo{author}{\bibfnamefont{J.~E.} \bibnamefont{Rhoads}},
  \bibinfo{author}{\bibfnamefont{S.}~\bibnamefont{Malhotra}},
  \bibinfo{author}{\bibfnamefont{R.}~\bibnamefont{Probst}},
  \bibinfo{author}{\bibfnamefont{R.}~\bibnamefont{Swaters}},
  \bibinfo{author}{\bibfnamefont{V.~S.} \bibnamefont{Tilvi}},
  \bibinfo{author}{\bibfnamefont{Z.-Y.} \bibnamefont{Zheng}},
  \bibinfo{author}{\bibfnamefont{S.}~\bibnamefont{Finkelstein}},
  \bibinfo{author}{\bibfnamefont{P.}~\bibnamefont{Hibon}},
  \bibinfo{author}{\bibfnamefont{B.}~\bibnamefont{Mobasher}},
  \bibnamefont{et~al.}, \bibinfo{journal}{The Astrophysical Journal}
  \textbf{\bibinfo{volume}{858}}, \bibinfo{pages}{96} (\bibinfo{year}{2018}),
  \urlprefix\url{https://doi.org/10.3847%2F1538-4357%2Faab620}.

\bibitem[{\citenamefont{{Khostovan} et~al.}(2020)\citenamefont{{Khostovan},
  {Malhotra}, {Rhoads}, {Jiang}, {Wang}, {Wold}, {Zheng}, {Barrientos},
  {Coughlin}, {Harish} et~al.}}]{2020MNRAS.tmp..171K}
\bibinfo{author}{\bibfnamefont{A.~A.} \bibnamefont{{Khostovan}}},
  \bibinfo{author}{\bibfnamefont{S.}~\bibnamefont{{Malhotra}}},
  \bibinfo{author}{\bibfnamefont{J.~E.} \bibnamefont{{Rhoads}}},
  \bibinfo{author}{\bibfnamefont{C.}~\bibnamefont{{Jiang}}},
  \bibinfo{author}{\bibfnamefont{J.}~\bibnamefont{{Wang}}},
  \bibinfo{author}{\bibfnamefont{I.}~\bibnamefont{{Wold}}},
  \bibinfo{author}{\bibfnamefont{Z.~Y.} \bibnamefont{{Zheng}}},
  \bibinfo{author}{\bibfnamefont{L.~F.} \bibnamefont{{Barrientos}}},
  \bibinfo{author}{\bibfnamefont{A.}~\bibnamefont{{Coughlin}}},
  \bibinfo{author}{\bibfnamefont{S.}~\bibnamefont{{Harish}}},
  \bibnamefont{et~al.}, \bibinfo{journal}{\mnras} p. \bibinfo{pages}{171}
  (\bibinfo{year}{2020}), \eprint{2001.04989}.

\bibitem[{\citenamefont{{Kennicutt}}(1998)}]{1998ARA&A..36..189K}
\bibinfo{author}{\bibfnamefont{J.}~\bibnamefont{{Kennicutt}},
  \bibfnamefont{Robert~C.}}, \bibinfo{journal}{\araa}
  \textbf{\bibinfo{volume}{36}}, \bibinfo{pages}{189} (\bibinfo{year}{1998}),
  \eprint{astro-ph/9807187}.

\bibitem[{\citenamefont{{Horiuchi} et~al.}(2009)\citenamefont{{Horiuchi},
  {Beacom}, and {Dwek}}}]{2009PhRvD..79h3013H}
\bibinfo{author}{\bibfnamefont{S.}~\bibnamefont{{Horiuchi}}},
  \bibinfo{author}{\bibfnamefont{J.~F.} \bibnamefont{{Beacom}}},
  \bibnamefont{and} \bibinfo{author}{\bibfnamefont{E.}~\bibnamefont{{Dwek}}},
  \bibinfo{journal}{\prd} \textbf{\bibinfo{volume}{79}}, \bibinfo{eid}{083013}
  (\bibinfo{year}{2009}), \eprint{0812.3157}.

\bibitem[{\citenamefont{Lee et~al.}(2009)}]{Lee:2009by}
\bibinfo{author}{\bibfnamefont{J.~C.} \bibnamefont{Lee}} \bibnamefont{et~al.},
  \bibinfo{journal}{Astrophys. J.} \textbf{\bibinfo{volume}{706}},
  \bibinfo{pages}{599} (\bibinfo{year}{2009}), \eprint{0909.5205}.

\bibitem[{\citenamefont{{Behroozi} et~al.}(2013)\citenamefont{{Behroozi},
  {Wechsler}, and {Conroy}}}]{2013ApJ...770...57B}
\bibinfo{author}{\bibfnamefont{P.~S.} \bibnamefont{{Behroozi}}},
  \bibinfo{author}{\bibfnamefont{R.~H.} \bibnamefont{{Wechsler}}},
  \bibnamefont{and} \bibinfo{author}{\bibfnamefont{C.}~\bibnamefont{{Conroy}}},
  \bibinfo{journal}{\apj} \textbf{\bibinfo{volume}{770}}, \bibinfo{eid}{57}
  (\bibinfo{year}{2013}), \eprint{1207.6105}.

\bibitem[{\citenamefont{{Wilkins} et~al.}(2019)\citenamefont{{Wilkins},
  {Lovell}, and {Stanway}}}]{2019MNRAS.490.5359W}
\bibinfo{author}{\bibfnamefont{S.~M.} \bibnamefont{{Wilkins}}},
  \bibinfo{author}{\bibfnamefont{C.~C.} \bibnamefont{{Lovell}}},
  \bibnamefont{and} \bibinfo{author}{\bibfnamefont{E.~R.}
  \bibnamefont{{Stanway}}}, \bibinfo{journal}{\mnras}
  \textbf{\bibinfo{volume}{490}}, \bibinfo{pages}{5359} (\bibinfo{year}{2019}),
  \eprint{1910.05220}.

\bibitem[{\citenamefont{{Mauch} et~al.}(2020)\citenamefont{{Mauch}, {Cotton},
  {Condon}, {Matthews}, {Abbott}, {Adam}, {Aldera}, {Asad}, {Bauermeister},
  {Bennett} et~al.}}]{2020ApJ...888...61M}
\bibinfo{author}{\bibfnamefont{T.}~\bibnamefont{{Mauch}}},
  \bibinfo{author}{\bibfnamefont{W.~D.} \bibnamefont{{Cotton}}},
  \bibinfo{author}{\bibfnamefont{J.~J.} \bibnamefont{{Condon}}},
  \bibinfo{author}{\bibfnamefont{A.~M.} \bibnamefont{{Matthews}}},
  \bibinfo{author}{\bibfnamefont{T.~D.} \bibnamefont{{Abbott}}},
  \bibinfo{author}{\bibfnamefont{R.~M.} \bibnamefont{{Adam}}},
  \bibinfo{author}{\bibfnamefont{M.~A.} \bibnamefont{{Aldera}}},
  \bibinfo{author}{\bibfnamefont{K.~M.~B.} \bibnamefont{{Asad}}},
  \bibinfo{author}{\bibfnamefont{E.~F.} \bibnamefont{{Bauermeister}}},
  \bibinfo{author}{\bibfnamefont{T.~G.~H.} \bibnamefont{{Bennett}}},
  \bibnamefont{et~al.}, \bibinfo{journal}{\apj} \textbf{\bibinfo{volume}{888}},
  \bibinfo{eid}{61} (\bibinfo{year}{2020}), \eprint{1912.06212}.

\bibitem[{\citenamefont{{Horiuchi} et~al.}(2011)\citenamefont{{Horiuchi},
  {Beacom}, {Kochanek}, {Prieto}, {Stanek}, and
  {Thompson}}}]{2011ApJ...738..154H}
\bibinfo{author}{\bibfnamefont{S.}~\bibnamefont{{Horiuchi}}},
  \bibinfo{author}{\bibfnamefont{J.~F.} \bibnamefont{{Beacom}}},
  \bibinfo{author}{\bibfnamefont{C.~S.} \bibnamefont{{Kochanek}}},
  \bibinfo{author}{\bibfnamefont{J.~L.} \bibnamefont{{Prieto}}},
  \bibinfo{author}{\bibfnamefont{K.~Z.} \bibnamefont{{Stanek}}},
  \bibnamefont{and} \bibinfo{author}{\bibfnamefont{T.~A.}
  \bibnamefont{{Thompson}}}, \bibinfo{journal}{\apj}
  \textbf{\bibinfo{volume}{738}}, \bibinfo{eid}{154} (\bibinfo{year}{2011}),
  \eprint{1102.1977}.

\bibitem[{\citenamefont{{Dahlen} et~al.}(2012)\citenamefont{{Dahlen},
  {Strolger}, {Riess}, {Mattila}, {Kankare}, and
  {Mobasher}}}]{2012ApJ...757...70D}
\bibinfo{author}{\bibfnamefont{T.}~\bibnamefont{{Dahlen}}},
  \bibinfo{author}{\bibfnamefont{L.-G.} \bibnamefont{{Strolger}}},
  \bibinfo{author}{\bibfnamefont{A.~G.} \bibnamefont{{Riess}}},
  \bibinfo{author}{\bibfnamefont{S.}~\bibnamefont{{Mattila}}},
  \bibinfo{author}{\bibfnamefont{E.}~\bibnamefont{{Kankare}}},
  \bibnamefont{and}
  \bibinfo{author}{\bibfnamefont{B.}~\bibnamefont{{Mobasher}}},
  \bibinfo{journal}{\apj} \textbf{\bibinfo{volume}{757}}, \bibinfo{eid}{70}
  (\bibinfo{year}{2012}), \eprint{1208.0342}.

\bibitem[{\citenamefont{Horiuchi et~al.}(2011)\citenamefont{Horiuchi, Beacom,
  Kochanek, Prieto, Stanek, and Thompson}}]{Horiuchi:2011zz}
\bibinfo{author}{\bibfnamefont{S.}~\bibnamefont{Horiuchi}},
  \bibinfo{author}{\bibfnamefont{J.~F.} \bibnamefont{Beacom}},
  \bibinfo{author}{\bibfnamefont{C.~S.} \bibnamefont{Kochanek}},
  \bibinfo{author}{\bibfnamefont{J.~L.} \bibnamefont{Prieto}},
  \bibinfo{author}{\bibfnamefont{K.~Z.} \bibnamefont{Stanek}},
  \bibnamefont{and} \bibinfo{author}{\bibfnamefont{T.~A.}
  \bibnamefont{Thompson}}, \bibinfo{journal}{Astrophys. J.}
  \textbf{\bibinfo{volume}{738}}, \bibinfo{pages}{154} (\bibinfo{year}{2011}),
  \eprint{1102.1977}.

\bibitem[{\citenamefont{Lien et~al.}(2010)\citenamefont{Lien, Fields, and
  Beacom}}]{Lien:2010yb}
\bibinfo{author}{\bibfnamefont{A.}~\bibnamefont{Lien}},
  \bibinfo{author}{\bibfnamefont{B.~D.} \bibnamefont{Fields}},
  \bibnamefont{and} \bibinfo{author}{\bibfnamefont{J.~F.}
  \bibnamefont{Beacom}}, \bibinfo{journal}{Phys. Rev.}
  \textbf{\bibinfo{volume}{D81}}, \bibinfo{pages}{083001}
  (\bibinfo{year}{2010}), \eprint{1001.3678}.

\bibitem[{\citenamefont{Bays et~al.}(2012)}]{Bays:2011si}
\bibinfo{author}{\bibfnamefont{K.}~\bibnamefont{Bays}} \bibnamefont{et~al.}
  (\bibinfo{collaboration}{Super-Kamiokande}), \bibinfo{journal}{Phys. Rev.}
  \textbf{\bibinfo{volume}{D85}}, \bibinfo{pages}{052007}
  (\bibinfo{year}{2012}), \eprint{1111.5031}.

\bibitem[{\citenamefont{Abe et~al.}(2018{\natexlab{a}})}]{Abe:2018uyc}
\bibinfo{author}{\bibfnamefont{K.}~\bibnamefont{Abe}} \bibnamefont{et~al.}
  (\bibinfo{collaboration}{Hyper-Kamiokande})
  (\bibinfo{year}{2018}{\natexlab{a}}), \eprint{1805.04163}.

\bibitem[{\citenamefont{Horiuchi et~al.}(2009)\citenamefont{Horiuchi, Beacom,
  and Dwek}}]{Horiuchi:2008jz}
\bibinfo{author}{\bibfnamefont{S.}~\bibnamefont{Horiuchi}},
  \bibinfo{author}{\bibfnamefont{J.~F.} \bibnamefont{Beacom}},
  \bibnamefont{and} \bibinfo{author}{\bibfnamefont{E.}~\bibnamefont{Dwek}},
  \bibinfo{journal}{Phys. Rev.} \textbf{\bibinfo{volume}{D79}},
  \bibinfo{pages}{083013} (\bibinfo{year}{2009}), \eprint{0812.3157}.

\bibitem[{\citenamefont{Sekiya}(2020)}]{Sekiya:2020pun}
\bibinfo{author}{\bibfnamefont{H.}~\bibnamefont{Sekiya}}, \bibinfo{journal}{J.\
  Phys.\ Conf.\ Ser.} \textbf{\bibinfo{volume}{1342}}, \bibinfo{pages}{012044}
  (\bibinfo{year}{2020}).

\bibitem[{\citenamefont{Beacom}(2010)}]{Beacom:2010kk}
\bibinfo{author}{\bibfnamefont{J.~F.} \bibnamefont{Beacom}},
  \bibinfo{journal}{Ann. Rev. Nucl. Part. Sci.} \textbf{\bibinfo{volume}{60}},
  \bibinfo{pages}{439} (\bibinfo{year}{2010}), \eprint{1004.3311}.

\bibitem[{\citenamefont{Hirata et~al.}(1987)\citenamefont{Hirata, Kajita,
  Koshiba, Nakahata, Oyama, Sato, Suzuki, Takita, Totsuka, Kifune
  et~al.}}]{PhysRevLett.58.1490}
\bibinfo{author}{\bibfnamefont{K.}~\bibnamefont{Hirata}},
  \bibinfo{author}{\bibfnamefont{T.}~\bibnamefont{Kajita}},
  \bibinfo{author}{\bibfnamefont{M.}~\bibnamefont{Koshiba}},
  \bibinfo{author}{\bibfnamefont{M.}~\bibnamefont{Nakahata}},
  \bibinfo{author}{\bibfnamefont{Y.}~\bibnamefont{Oyama}},
  \bibinfo{author}{\bibfnamefont{N.}~\bibnamefont{Sato}},
  \bibinfo{author}{\bibfnamefont{A.}~\bibnamefont{Suzuki}},
  \bibinfo{author}{\bibfnamefont{M.}~\bibnamefont{Takita}},
  \bibinfo{author}{\bibfnamefont{Y.}~\bibnamefont{Totsuka}},
  \bibinfo{author}{\bibfnamefont{T.}~\bibnamefont{Kifune}},
  \bibnamefont{et~al.}, \bibinfo{journal}{Phys. Rev. Lett.}
  \textbf{\bibinfo{volume}{58}}, \bibinfo{pages}{1490} (\bibinfo{year}{1987}),
  \urlprefix\url{https://link.aps.org/doi/10.1103/PhysRevLett.58.1490}.

\bibitem[{\citenamefont{Hirata et~al.}(1988)\citenamefont{Hirata, Kajita,
  Koshiba, Nakahata, Oyama, Sato, Suzuki, Takita, Totsuka, Kifune
  et~al.}}]{PhysRevD.38.448}
\bibinfo{author}{\bibfnamefont{K.~S.} \bibnamefont{Hirata}},
  \bibinfo{author}{\bibfnamefont{T.}~\bibnamefont{Kajita}},
  \bibinfo{author}{\bibfnamefont{M.}~\bibnamefont{Koshiba}},
  \bibinfo{author}{\bibfnamefont{M.}~\bibnamefont{Nakahata}},
  \bibinfo{author}{\bibfnamefont{Y.}~\bibnamefont{Oyama}},
  \bibinfo{author}{\bibfnamefont{N.}~\bibnamefont{Sato}},
  \bibinfo{author}{\bibfnamefont{A.}~\bibnamefont{Suzuki}},
  \bibinfo{author}{\bibfnamefont{M.}~\bibnamefont{Takita}},
  \bibinfo{author}{\bibfnamefont{Y.}~\bibnamefont{Totsuka}},
  \bibinfo{author}{\bibfnamefont{T.}~\bibnamefont{Kifune}},
  \bibnamefont{et~al.}, \bibinfo{journal}{Phys. Rev. D}
  \textbf{\bibinfo{volume}{38}}, \bibinfo{pages}{448} (\bibinfo{year}{1988}),
  \urlprefix\url{https://link.aps.org/doi/10.1103/PhysRevD.38.448}.

\bibitem[{\citenamefont{Bionta et~al.}(1987)\citenamefont{Bionta, Blewitt,
  Bratton, Casper, Ciocio, Claus, Cortez, Crouch, Dye, Errede
  et~al.}}]{PhysRevLett.58.1494}
\bibinfo{author}{\bibfnamefont{R.~M.} \bibnamefont{Bionta}},
  \bibinfo{author}{\bibfnamefont{G.}~\bibnamefont{Blewitt}},
  \bibinfo{author}{\bibfnamefont{C.~B.} \bibnamefont{Bratton}},
  \bibinfo{author}{\bibfnamefont{D.}~\bibnamefont{Casper}},
  \bibinfo{author}{\bibfnamefont{A.}~\bibnamefont{Ciocio}},
  \bibinfo{author}{\bibfnamefont{R.}~\bibnamefont{Claus}},
  \bibinfo{author}{\bibfnamefont{B.}~\bibnamefont{Cortez}},
  \bibinfo{author}{\bibfnamefont{M.}~\bibnamefont{Crouch}},
  \bibinfo{author}{\bibfnamefont{S.~T.} \bibnamefont{Dye}},
  \bibinfo{author}{\bibfnamefont{S.}~\bibnamefont{Errede}},
  \bibnamefont{et~al.}, \bibinfo{journal}{Phys. Rev. Lett.}
  \textbf{\bibinfo{volume}{58}}, \bibinfo{pages}{1494} (\bibinfo{year}{1987}),
  \urlprefix\url{https://link.aps.org/doi/10.1103/PhysRevLett.58.1494}.

\bibitem[{\citenamefont{Bratton et~al.}(1988)\citenamefont{Bratton, Casper,
  Ciocio, Claus, Crouch, Dye, Errede, Gajewski, Goldhaber, Haines
  et~al.}}]{PhysRevD.37.3361}
\bibinfo{author}{\bibfnamefont{C.~B.} \bibnamefont{Bratton}},
  \bibinfo{author}{\bibfnamefont{D.}~\bibnamefont{Casper}},
  \bibinfo{author}{\bibfnamefont{A.}~\bibnamefont{Ciocio}},
  \bibinfo{author}{\bibfnamefont{R.}~\bibnamefont{Claus}},
  \bibinfo{author}{\bibfnamefont{M.}~\bibnamefont{Crouch}},
  \bibinfo{author}{\bibfnamefont{S.~T.} \bibnamefont{Dye}},
  \bibinfo{author}{\bibfnamefont{S.}~\bibnamefont{Errede}},
  \bibinfo{author}{\bibfnamefont{W.}~\bibnamefont{Gajewski}},
  \bibinfo{author}{\bibfnamefont{M.}~\bibnamefont{Goldhaber}},
  \bibinfo{author}{\bibfnamefont{T.~J.} \bibnamefont{Haines}},
  \bibnamefont{et~al.}, \bibinfo{journal}{Phys. Rev. D}
  \textbf{\bibinfo{volume}{37}}, \bibinfo{pages}{3361} (\bibinfo{year}{1988}),
  \urlprefix\url{https://link.aps.org/doi/10.1103/PhysRevD.37.3361}.

\bibitem[{\citenamefont{{Alekseev} et~al.}(1987)\citenamefont{{Alekseev},
  {Alekseeva}, {Volchenko}, and {Krivosheina}}}]{1987JETPL..45..589A}
\bibinfo{author}{\bibfnamefont{E.~N.} \bibnamefont{{Alekseev}}},
  \bibinfo{author}{\bibfnamefont{L.~N.} \bibnamefont{{Alekseeva}}},
  \bibinfo{author}{\bibfnamefont{V.~I.} \bibnamefont{{Volchenko}}},
  \bibnamefont{and} \bibinfo{author}{\bibfnamefont{I.~V.}
  \bibnamefont{{Krivosheina}}}, \bibinfo{journal}{Soviet Journal of
  Experimental and Theoretical Physics Letters} \textbf{\bibinfo{volume}{45}},
  \bibinfo{pages}{589} (\bibinfo{year}{1987}).

\bibitem[{\citenamefont{Ikeda et~al.}(2007)\citenamefont{Ikeda, Takeda, Fukuda,
  Vagins, Abe, Iida, Ishihara, Kameda, Koshio, Minamino et~al.}}]{Ikeda_2007}
\bibinfo{author}{\bibfnamefont{M.}~\bibnamefont{Ikeda}},
  \bibinfo{author}{\bibfnamefont{A.}~\bibnamefont{Takeda}},
  \bibinfo{author}{\bibfnamefont{Y.}~\bibnamefont{Fukuda}},
  \bibinfo{author}{\bibfnamefont{M.~R.} \bibnamefont{Vagins}},
  \bibinfo{author}{\bibfnamefont{K.}~\bibnamefont{Abe}},
  \bibinfo{author}{\bibfnamefont{T.}~\bibnamefont{Iida}},
  \bibinfo{author}{\bibfnamefont{K.}~\bibnamefont{Ishihara}},
  \bibinfo{author}{\bibfnamefont{J.}~\bibnamefont{Kameda}},
  \bibinfo{author}{\bibfnamefont{Y.}~\bibnamefont{Koshio}},
  \bibinfo{author}{\bibfnamefont{A.}~\bibnamefont{Minamino}},
  \bibnamefont{et~al.}, \bibinfo{journal}{The Astrophysical Journal}
  \textbf{\bibinfo{volume}{669}}, \bibinfo{pages}{519} (\bibinfo{year}{2007}),
  \urlprefix\url{https://doi.org/10.1086%2F521547}.

\bibitem[{\citenamefont{Migenda}(2018)}]{Migenda:2018ljh}
\bibinfo{author}{\bibfnamefont{J.}~\bibnamefont{Migenda}}
  (\bibinfo{collaboration}{DUNE}), in \emph{\bibinfo{booktitle}{{Prospects in
  Neutrino Physics}}} (\bibinfo{year}{2018}), pp. \bibinfo{pages}{164--168},
  \eprint{1804.01877}.

\bibitem[{\citenamefont{{Tammann} et~al.}(1994)\citenamefont{{Tammann},
  {Loeffler}, and {Schroeder}}}]{1994ApJS...92..487T}
\bibinfo{author}{\bibfnamefont{G.~A.} \bibnamefont{{Tammann}}},
  \bibinfo{author}{\bibfnamefont{W.}~\bibnamefont{{Loeffler}}},
  \bibnamefont{and}
  \bibinfo{author}{\bibfnamefont{A.}~\bibnamefont{{Schroeder}}},
  \bibinfo{journal}{\apjs} \textbf{\bibinfo{volume}{92}}, \bibinfo{pages}{487}
  (\bibinfo{year}{1994}).

\bibitem[{\citenamefont{Ando et~al.}(2005{\natexlab{a}})\citenamefont{Ando,
  Beacom, and Y\"uksel}}]{PhysRevLett.95.171101}
\bibinfo{author}{\bibfnamefont{S.}~\bibnamefont{Ando}},
  \bibinfo{author}{\bibfnamefont{J.~F.} \bibnamefont{Beacom}},
  \bibnamefont{and} \bibinfo{author}{\bibfnamefont{H.}~\bibnamefont{Y\"uksel}},
  \bibinfo{journal}{Phys. Rev. Lett.} \textbf{\bibinfo{volume}{95}},
  \bibinfo{pages}{171101} (\bibinfo{year}{2005}{\natexlab{a}}),
  \urlprefix\url{https://link.aps.org/doi/10.1103/PhysRevLett.95.171101}.

\bibitem[{\citenamefont{{Nakamura}}(2003)}]{2003IJMPA..18.4053N}
\bibinfo{author}{\bibfnamefont{K.}~\bibnamefont{{Nakamura}}},
  \bibinfo{journal}{International Journal of Modern Physics A}
  \textbf{\bibinfo{volume}{18}}, \bibinfo{pages}{4053} (\bibinfo{year}{2003}).

\bibitem[{\citenamefont{Jung}(2000)}]{Jung:1999jq}
\bibinfo{author}{\bibfnamefont{C.~K.} \bibnamefont{Jung}},
  \bibinfo{journal}{AIP Conf. Proc.} \textbf{\bibinfo{volume}{533}},
  \bibinfo{pages}{29} (\bibinfo{year}{2000}), \eprint{hep-ex/0005046}.

\bibitem[{\citenamefont{de~Bellefon et~al.}(2006)}]{deBellefon:2006vq}
\bibinfo{author}{\bibfnamefont{A.}~\bibnamefont{de~Bellefon}}
  \bibnamefont{et~al.} (\bibinfo{year}{2006}), \eprint{hep-ex/0607026}.

\bibitem[{\citenamefont{Agostino et~al.}(2013)\citenamefont{Agostino,
  Buizza-Avanzini, Dracos, Duchesneau, Marafini, Mezzetto, Mosca, Patzak,
  Tonazzo, and Vassilopoulos}}]{Agostino:2012fd}
\bibinfo{author}{\bibfnamefont{L.}~\bibnamefont{Agostino}},
  \bibinfo{author}{\bibfnamefont{M.}~\bibnamefont{Buizza-Avanzini}},
  \bibinfo{author}{\bibfnamefont{M.}~\bibnamefont{Dracos}},
  \bibinfo{author}{\bibfnamefont{D.}~\bibnamefont{Duchesneau}},
  \bibinfo{author}{\bibfnamefont{M.}~\bibnamefont{Marafini}},
  \bibinfo{author}{\bibfnamefont{M.}~\bibnamefont{Mezzetto}},
  \bibinfo{author}{\bibfnamefont{L.}~\bibnamefont{Mosca}},
  \bibinfo{author}{\bibfnamefont{T.}~\bibnamefont{Patzak}},
  \bibinfo{author}{\bibfnamefont{A.}~\bibnamefont{Tonazzo}}, \bibnamefont{and}
  \bibinfo{author}{\bibfnamefont{N.}~\bibnamefont{Vassilopoulos}}
  (\bibinfo{collaboration}{MEMPHYS}), \bibinfo{journal}{JCAP}
  \textbf{\bibinfo{volume}{01}}, \bibinfo{pages}{024} (\bibinfo{year}{2013}),
  \eprint{1206.6665}.

\bibitem[{\citenamefont{Pattavina et~al.}(2020)\citenamefont{Pattavina,
  Iachellini, and Tamborra}}]{Pattavina:2020cqc}
\bibinfo{author}{\bibfnamefont{L.}~\bibnamefont{Pattavina}},
  \bibinfo{author}{\bibfnamefont{N.~F.} \bibnamefont{Iachellini}},
  \bibnamefont{and} \bibinfo{author}{\bibfnamefont{I.}~\bibnamefont{Tamborra}}
  (\bibinfo{year}{2020}), \eprint{2004.06936}.

\bibitem[{\citenamefont{Suzuki et~al.}(2001)}]{Suzuki:2001rb}
\bibinfo{author}{\bibfnamefont{Y.}~\bibnamefont{Suzuki}} \bibnamefont{et~al.}
  (\bibinfo{collaboration}{TITAND Working Group}), in
  \emph{\bibinfo{booktitle}{{11th International School on Particles and
  Cosmology Karbardino-Balkaria, Russia, April 18-24, 2001}}}
  (\bibinfo{year}{2001}), pp. \bibinfo{pages}{288--296},
  \bibinfo{note}{[,288(2001)]}, \eprint{hep-ex/0110005}.

\bibitem[{\citenamefont{Böser et~al.}(2015)\citenamefont{Böser, Kowalski,
  Schulte, Strotjohann, and Voge}}]{Boser:2013oaa}
\bibinfo{author}{\bibfnamefont{S.}~\bibnamefont{Böser}},
  \bibinfo{author}{\bibfnamefont{M.}~\bibnamefont{Kowalski}},
  \bibinfo{author}{\bibfnamefont{L.}~\bibnamefont{Schulte}},
  \bibinfo{author}{\bibfnamefont{N.~L.} \bibnamefont{Strotjohann}},
  \bibnamefont{and} \bibinfo{author}{\bibfnamefont{M.}~\bibnamefont{Voge}},
  \bibinfo{journal}{Astropart. Phys.} \textbf{\bibinfo{volume}{62}},
  \bibinfo{pages}{54} (\bibinfo{year}{2015}), \eprint{1304.2553}.

\bibitem[{\citenamefont{Kistler et~al.}(2011)\citenamefont{Kistler, Yüksel,
  Ando, Beacom, and Suzuki}}]{Kistler_2011}
\bibinfo{author}{\bibfnamefont{M.~D.} \bibnamefont{Kistler}},
  \bibinfo{author}{\bibfnamefont{H.}~\bibnamefont{Yüksel}},
  \bibinfo{author}{\bibfnamefont{S.}~\bibnamefont{Ando}},
  \bibinfo{author}{\bibfnamefont{J.~F.} \bibnamefont{Beacom}},
  \bibnamefont{and} \bibinfo{author}{\bibfnamefont{Y.}~\bibnamefont{Suzuki}},
  \bibinfo{journal}{Physical Review D} \textbf{\bibinfo{volume}{83}}
  (\bibinfo{year}{2011}), ISSN \bibinfo{issn}{1550-2368},
  \urlprefix\url{http://dx.doi.org/10.1103/PhysRevD.83.123008}.

\bibitem[{\citenamefont{Ando et~al.}(2005{\natexlab{b}})\citenamefont{Ando,
  Beacom, and Yuksel}}]{Ando:2005ka}
\bibinfo{author}{\bibfnamefont{S.}~\bibnamefont{Ando}},
  \bibinfo{author}{\bibfnamefont{J.~F.} \bibnamefont{Beacom}},
  \bibnamefont{and} \bibinfo{author}{\bibfnamefont{H.}~\bibnamefont{Yuksel}},
  \bibinfo{journal}{Phys. Rev. Lett.} \textbf{\bibinfo{volume}{95}},
  \bibinfo{pages}{171101} (\bibinfo{year}{2005}{\natexlab{b}}),
  \eprint{astro-ph/0503321}.

\bibitem[{\citenamefont{Yuksel et~al.}(2006)\citenamefont{Yuksel, Ando, and
  Beacom}}]{Yuksel:2005ae}
\bibinfo{author}{\bibfnamefont{H.}~\bibnamefont{Yuksel}},
  \bibinfo{author}{\bibfnamefont{S.}~\bibnamefont{Ando}}, \bibnamefont{and}
  \bibinfo{author}{\bibfnamefont{J.~F.} \bibnamefont{Beacom}},
  \bibinfo{journal}{Phys. Rev.} \textbf{\bibinfo{volume}{C74}},
  \bibinfo{pages}{015803} (\bibinfo{year}{2006}), \eprint{astro-ph/0509297}.

\bibitem[{\citenamefont{Lunardini and Tamborra}(2012)}]{Lunardini:2012ne}
\bibinfo{author}{\bibfnamefont{C.}~\bibnamefont{Lunardini}} \bibnamefont{and}
  \bibinfo{author}{\bibfnamefont{I.}~\bibnamefont{Tamborra}},
  \bibinfo{journal}{JCAP} \textbf{\bibinfo{volume}{1207}}, \bibinfo{pages}{012}
  (\bibinfo{year}{2012}), \eprint{1205.6292}.

\bibitem[{\citenamefont{Adams et~al.}(2013)\citenamefont{Adams, Kochanek,
  Beacom, Vagins, and Stanek}}]{Adams:2013ana}
\bibinfo{author}{\bibfnamefont{S.~M.} \bibnamefont{Adams}},
  \bibinfo{author}{\bibfnamefont{C.~S.} \bibnamefont{Kochanek}},
  \bibinfo{author}{\bibfnamefont{J.~F.} \bibnamefont{Beacom}},
  \bibinfo{author}{\bibfnamefont{M.~R.} \bibnamefont{Vagins}},
  \bibnamefont{and} \bibinfo{author}{\bibfnamefont{K.~Z.}
  \bibnamefont{Stanek}}, \bibinfo{journal}{Astrophys. J.}
  \textbf{\bibinfo{volume}{778}}, \bibinfo{pages}{164} (\bibinfo{year}{2013}),
  \eprint{1306.0559}.

\bibitem[{\citenamefont{Nikrant et~al.}(2018)\citenamefont{Nikrant, Laha, and
  Horiuchi}}]{Nikrant:2017nya}
\bibinfo{author}{\bibfnamefont{A.}~\bibnamefont{Nikrant}},
  \bibinfo{author}{\bibfnamefont{R.}~\bibnamefont{Laha}}, \bibnamefont{and}
  \bibinfo{author}{\bibfnamefont{S.}~\bibnamefont{Horiuchi}},
  \bibinfo{journal}{Phys. Rev.} \textbf{\bibinfo{volume}{D97}},
  \bibinfo{pages}{023019} (\bibinfo{year}{2018}), \eprint{1711.00008}.

\bibitem[{\citenamefont{Horiuchi and Kneller}(2018)}]{Horiuchi:2017sku}
\bibinfo{author}{\bibfnamefont{S.}~\bibnamefont{Horiuchi}} \bibnamefont{and}
  \bibinfo{author}{\bibfnamefont{J.~P.} \bibnamefont{Kneller}},
  \bibinfo{journal}{J. Phys.} \textbf{\bibinfo{volume}{G45}},
  \bibinfo{pages}{043002} (\bibinfo{year}{2018}), \eprint{1709.01515}.

\bibitem[{\citenamefont{Migenda}(2019)}]{Migenda:2020rot}
\bibinfo{author}{\bibfnamefont{J.}~\bibnamefont{Migenda}}, Ph.D. thesis,
  \bibinfo{school}{Sheffield U.} (\bibinfo{year}{2019}), \eprint{2002.01649}.

\bibitem[{\citenamefont{Yoshida et~al.}(2005)\citenamefont{Yoshida, Kajino, and
  Hartmann}}]{Yoshida:2005uy}
\bibinfo{author}{\bibfnamefont{T.}~\bibnamefont{Yoshida}},
  \bibinfo{author}{\bibfnamefont{T.}~\bibnamefont{Kajino}}, \bibnamefont{and}
  \bibinfo{author}{\bibfnamefont{D.~H.} \bibnamefont{Hartmann}},
  \bibinfo{journal}{Phys.\ Rev.\ Lett.} \textbf{\bibinfo{volume}{94}},
  \bibinfo{pages}{231101} (\bibinfo{year}{2005}), \eprint{astro-ph/0505043}.

\bibitem[{\citenamefont{Lunardini}(2009)}]{Lunardini:2009ya}
\bibinfo{author}{\bibfnamefont{C.}~\bibnamefont{Lunardini}},
  \bibinfo{journal}{Phys. Rev. Lett.} \textbf{\bibinfo{volume}{102}},
  \bibinfo{pages}{231101} (\bibinfo{year}{2009}), \eprint{0901.0568}.

\bibitem[{\citenamefont{Tamborra}(2016)}]{Tamborra:2016lpf}
\bibinfo{author}{\bibfnamefont{I.}~\bibnamefont{Tamborra}}, in
  \emph{\bibinfo{booktitle}{{Prospects in Neutrino Physics}}}
  (\bibinfo{year}{2016}), \eprint{1604.07332}.

\bibitem[{\citenamefont{Priya and Lunardini}(2017)}]{Priya:2017bmm}
\bibinfo{author}{\bibfnamefont{A.}~\bibnamefont{Priya}} \bibnamefont{and}
  \bibinfo{author}{\bibfnamefont{C.}~\bibnamefont{Lunardini}},
  \bibinfo{journal}{JCAP} \textbf{\bibinfo{volume}{1711}}, \bibinfo{pages}{031}
  (\bibinfo{year}{2017}), \eprint{1705.02122}.

\bibitem[{\citenamefont{Abe et~al.}(2018{\natexlab{b}})}]{Abe:2016ero}
\bibinfo{author}{\bibfnamefont{K.}~\bibnamefont{Abe}} \bibnamefont{et~al.}
  (\bibinfo{collaboration}{Hyper-Kamiokande}), \bibinfo{journal}{PTEP}
  \textbf{\bibinfo{volume}{2018}}, \bibinfo{pages}{063C01}
  (\bibinfo{year}{2018}{\natexlab{b}}), \eprint{1611.06118}.

\bibitem[{\citenamefont{Lunardini}(2016)}]{Lunardini:2010ab}
\bibinfo{author}{\bibfnamefont{C.}~\bibnamefont{Lunardini}},
  \bibinfo{journal}{Astropart. Phys.} \textbf{\bibinfo{volume}{79}},
  \bibinfo{pages}{49} (\bibinfo{year}{2016}), \eprint{1007.3252}.

\bibitem[{\citenamefont{Beacom and Vagins}(2004)}]{Beacom:2003nk}
\bibinfo{author}{\bibfnamefont{J.~F.} \bibnamefont{Beacom}} \bibnamefont{and}
  \bibinfo{author}{\bibfnamefont{M.~R.} \bibnamefont{Vagins}},
  \bibinfo{journal}{Phys. Rev. Lett.} \textbf{\bibinfo{volume}{93}},
  \bibinfo{pages}{171101} (\bibinfo{year}{2004}), \eprint{hep-ph/0309300}.

\bibitem[{\citenamefont{Bays}(2012)}]{Bays:2012wty}
\bibinfo{author}{\bibfnamefont{K.~R.} \bibnamefont{Bays}}, Ph.D. thesis,
  \bibinfo{school}{UC, Irvine} (\bibinfo{year}{2012}).

\bibitem[{\citenamefont{Zhang et~al.}(1988)}]{Zhang:1988tv}
\bibinfo{author}{\bibfnamefont{W.}~\bibnamefont{Zhang}} \bibnamefont{et~al.},
  \bibinfo{journal}{Phys.\ Rev.\ Lett.} \textbf{\bibinfo{volume}{61}},
  \bibinfo{pages}{385} (\bibinfo{year}{1988}).

\end{thebibliography}
\end{document}